# An Adaptive Estimation Approach based on Fisher Information to Overcome the Challenges of LFP Battery SOC Estimation

Junzhe Shi[1], Shida Jiang[2], Shengyu Tao[3]*, Jaewong Lee[4], Manashita Borah[5], Scott Moura[6]

1. University of California, Berkeley, Department of Civil and Environmental Engineering, 760 Davis Hall, Berkeley, CA 94720, USA. Berkeley, CA 94720, USA. junzhe@berkeley.edu
2. University of California, Berkeley, Department of Civil and Environmental Engineering, 760 Davis Hall, Berkeley, CA 94720, USA. Berkeley, CA 94720, USA. shida_jiang@berkeley.edu
3. University of California, Berkeley, Department of Civil and Environmental Engineering, 760 Davis Hall, Berkeley, CA 94720, USA. Berkeley, CA 94720, USA. sytao@berkeley.edu
4. University of California, Berkeley, Department of Civil and Environmental Engineering, 760 Davis Hall, Berkeley, CA 94720, USA. Berkeley, CA 94720, USA. ljw7696@berkeley.edu
5. University of California, Berkeley, Department of Civil and Environmental Engineering, 760 Davis Hall, Berkeley, CA 94720, USA. Berkeley, CA 94720, USA. Tezpur University, Department of Electrical Engineering, Assam, 784028, India. manashitaborah@berkeley.edu
6. University of California, Berkeley, Department of Civil and Environmental Engineering, 760 Davis Hall, Berkeley, CA 94720, USA. Berkeley, CA 94720, USA. smoura@berkeley.edu

* Corresponding author: Shengyu Tao, University of California, Berkeley, Department of Civil and Environmental Engineering, 760 Davis Hall, Berkeley, CA 94720, USA, Email: sytao@berkeley.edu

*Abstract*— Robust and Real-time State of Charge (SOC) estimation is essential for Lithium Iron Phosphate (LFP) batteries, which are widely used in electric vehicles (EVs) and energy storage systems due to safety and longevity. However, the flat Open Circuit Voltage (OCV)-SOC curve makes this task particularly challenging. This challenge is complicated by hysteresis effects, and real-world conditions such as current bias, voltage quantization errors, and temperature that must be considered in the battery management system use. In this paper, we proposed an adaptive estimation approach to overcome the challenges of LFPSOC estimation. Specifically, the method uses an adaptive fisher information fusion strategy that adaptively combines the SOC estimation from two different models, which are Coulomb counting and equivalent circuit model-based parameter identification. The effectiveness of this strategy is rationalized by the information richness excited by external cycling signals. A 3D OCV-H-SOC map that captures the relationship between OCV, hysteresis, and SOC was proposed as the backbone, and can be generalizable to other widely adopted parameter-identification methods. Extensive validation under ideal and real-world use scenarios, including SOC-OCV flat zones, current bias, voltage quantization errors, low temperatures, and insufficient current excitations, have been performed using 4 driving profiles, i.e., the Orange County Transit Bus Cycle, the California Unified Cycle, the US06 Drive Cycle, and the New York City Cycle, where the results demonstrate superiority over the state-of-the-art unscented Kalman filter, long short-term memory networks and transformer in all validation cases.

*Table of Content Image*—

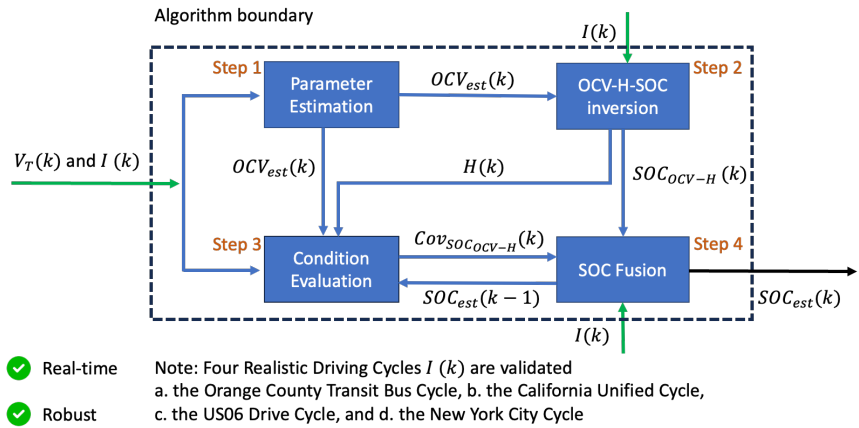

*Keywords*— *Li-ion batteries, SOC estimation, Battery management system, Adaptive control, Robust estimation*

I. INTRODUCTION

*A. Background and Motivation*

Lithium Iron Phosphate (LFP) batteries stand out for their safety, long cycle life, and cost-effectiveness in applications such as electric vehicles (EVs) and energy storage systems. The accuracy of SOC estimation directly impacts operational efficiency by preventing overcharging or deep discharging, thereby enhancing battery life and safety [1]. It facilitates precise energy management in EVs and storage systems [2], allowing for optimized energy allocation [3]. Moreover, reliable SOC estimation improves user confidence in EVs by providing accurate range predictions and optimal route planning [4]. These needs necessitate advanced, high accuracy SOC estimation techniques for LFP cells that can adapt to diverse operating conditions.

However, SOC estimation for LFP batteries presents unique challenges due to their flat SOC-Open Circuit Voltage (OCV) curve [5]. This flat SOC-OCV plateau weakens observability by mapping measured voltage signals to SOC value to estimate. Additionally, hysteresis effects in LFP cells significantly influence OCV variations, which can be more than SOC changes themselves. In real-world applications, non-ideal conditions such as operation in SOC-OCV flat zones, current bias, voltage quantization errors, and temperature variations further complex the SOC estimation challenge in the presence of hysteresis effects [6]. Thus, a reliable, efficient, and flexible hysteresis effects modeling techniques for LFP batteries are urgently needed.

*B. Literature review*

Accurate SOC estimation is crucial for the effective LFP battery management. The most basic method is Coulomb counting, often employed due to its simplicity. However, as an open-loop method, Coulomb counting suffers from two significant drawbacks: (1) the inability to correct for initial SOC errors and (2) the accumulation of errors due to current bias and capacity estimation inaccuracies [7] [8].

To address these issues, various advanced methods have been developed. Physics-based methods utilize mathematical models to describe the electrochemical processes and dynamic behavior of batteries. The motivation for using model-based methods lies in their ability to provide a physically interpretable framework that accurately represents the internal states and dynamics of the battery [9]. These methods employ equivalent circuit models (ECM) or more complex electrochemical models, combining Coulomb counting and the inverse of the cell's OCV to get a more accurate SOC estimation. The most popular model-based methods include the extended Kalman filter (EKF) [10] and unscented Kalman filter (UKF) [11]. The EKF is widely used due to its ability to handle non-linear dynamics and recognize model and measurement uncertainties [12]. In [13], an EKF observer is developed and demonstrated to have better performance than a Luenberger observer in LFP cell SOC estimation. However, the EKF suffers from linear approximation issues that compromises battery dynamic behaviors. The UKF addresses linearity limitations by using a deterministic sampling approach to capture mean and covariance estimates more accurately. A UKF applied to a reduced-order model for highly non-linear lithium-ion concentration and SOC estimation is presented in [14]. However, the performance of these model-based methods is highly sensitive to the model parameters [15]. As these parameters become less accurate with battery aging, temperature changes, and current variations, the performance of these methods deteriorates. Additionally, these methods typically model the measurement covariance matrix and process noise covariance matrix as constant hyperparameters, so the fusion of the Coulomb counting and the inverse of the OCV function is usually suboptimal and biased [16], [17].

Data-driven methods leverage machine learning and statistical analysis methodologies to predict SOC based on historical data and observed patterns [18]. The motivation of applying data-driven methods stems from their ability to model complex, non-linear relationships without requiring detailed first principles models [19]. The typical data-driven methods include deep neural networks (DNN), long short-term memory (LSTM) networks, and transformer models [20]. DNNs are capable of modeling complex, non-linear relationships between battery measurement data and SOC [21]. Since battery measurement data are time-series data, LSTM networks, a type of recurrent neural network (RNN), are suitable to capture long-term temporal dependencies in battery behavior. An LSTM-based approach for SOC estimation is developed in [22], showing a fast convergence speed to the true SOC with sequential current, voltage, and temperature measurement data as the inputs. To focus more on the entire input sequence and address the vanishing gradient problem, transformer models with attention mechanisms have been adapted in the battery state estimation field. A transformer model for SOC estimation was investigated in [23] with a 64-second sliding window size, yielding better results than the LSTM method. Despite their superior in capturing highly non-linear measurement-target pairs, interpretability remains a significant bottleneck [24]. More importantly, data-driven model performances are obtained under designed test conditions with specific training data [25] [26] [27]. While in practical scenarios such as battery reuse and recycling, these measurement data is scarce and heterogeneous, calling for more extensive data curation or more advances learning techniques such as collaborative and generative machine learning [28] [29] [30] [31].

Compared to well-studied NMC cells [32], accurate SOC estimation for LFP batteries remains more challenging due to their flat OCV characteristics and inappropriate hysteresis effects modeling. For instance, [33] proposes a NARX dynamical neural network to address the flat SOC-OCV curve of LFP cells but sacrifices time efficiency for accuracy. Similarly, [34] introduces an adaptive recursive square root algorithm for real-time OCV and parameter identification but overlooks hysteresis and operational condition impacts. Due to the flat OCV-SOC curve, the effect of hysteresis on OCV variations can be more significant than the SOC, thus neglecting hysteresis in LFP cell SOC estimation can lead to considerable errors. While Jöst et al. [35] demonstrate robust UKF performance under frequency containment reserve conditions, hysteresis-induced voltage ambiguities and current measurement errors degrade the estimation accuracy with dynamic load shifts. Weak SOC-OCV observability further exacerbates vulnerability to sensor biases, particularly when LFP cells operate in their flat OCV range for extended periods. Shi et al. [7] combine online parameter estimation with DNN to mitigate current bias but struggle under constant-current conditions due to insufficient excitation persistence. These efforts underscore unresolved challenges: (1) balancing robustness with computational efficiency, (2) ensuring adaptability to varying operational conditions, and (3) addressing the intertwined effects of flat SOC-OCV, hysteresis, and sensor bias.

In summary, current battery SOC estimation methods face several significant limitations. The mainstream approaches lack the adaptability needed to account for changes in battery characteristics, such as hysteresis effect, over time. While online parameter estimation methods offer the potential to update parameters, they perform poorly when persistency of excitation conditions are not satisfied due to the flatness of OCV-SOC relationship, and the hysteresis effect that further complicates this relationship. To the best of the authors' knowledge, no existing work offers an integrated approach that uses the information gain resulted from the dynamic cycling profiles to guide automated selection of SOC estimation, which is named as the SOC fusion, from classical coulomb counting and parameter-identification based method. This significant gap can lead to inaccuracies of SOC estimation in prolonged operation in the OCV-SOC flat zone and with high current bias, which is a common but outstanding challenge in both EV and grid energy storage systems.

*C. Contribution*

This work addresses the aforementioned research gaps by developing an adaptive and real-time SOC estimation approach for LFP batteries. Specifically, this paper enhances SOC estimation accuracy and reliability under diverse operational conditions, SOC-OCV flat zones, current bias, voltage quantization errors, and varying temperatures. To evaluate our method's effectiveness, we conduct a comprehensive comparison with several widely used state-of-charge estimation techniques, including unscented UKF, LSTM, and decoder-only Transformer model. We also implement hybrid fusion approaches that combine data-driven models with Coulomb counting via Kalman filtering. The proposed method consistently outperforms all benchmarks across five challenging test conditions, demonstrating superior accuracy, robustness, and computational efficiency suitable for real-time applications.

The key contributions of this study include:
- The novel SOC fusion strategy was proposed for adaptively combining Coulomb counting and parameter-identification based method using the information gain resulted from the dynamic cycling profiles.
- The first 3D map that captures the relationship between OCV, hysteresis, and SOC was proposed as the backbone of widely adopted parameter-identification based method, which now lacks of the critical hysteresis effect information.
- The proposed SOC estimation algorithm is mathematically provable that the estimation performance is irrelevant to current measurement bias, which can be regarded as the most significant error source of existing methods.

*D. Organization of the Paper*

The organization of this paper is as follows. Section II introduces the methodologies and the four different modules employed in this study. Section III presents the test results and provides an analysis of these results under various operational conditions. Lastly, Section IV encapsulates the key conclusions drawn from this study.

II. METHODOLOGY

This section provides an overview of the system. It clarifies the role and functionality of each module within our proposed framework for SOC estimation in LFP batteries. Following this overview, we delve into the specific methodologies of each module in the subsequent subsections.

## A. System Overview

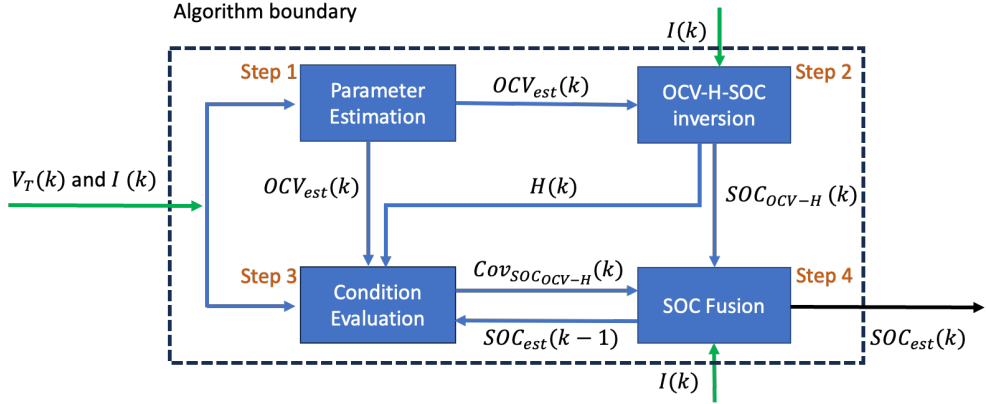

Fig. 1. Frame and flow chart of the proposed system.

Fig. 1 presents a schematic representation of the proposed system's architecture and flow. The system is structured into four distinct but interconnected modules, each contributing a vital element to the overall SOC estimation process. The model parameter structure can be found in the Table 1. A brief summary of each module is as follows:

1. **Parameter Estimation Module (PEM):** This module forms the foundation of our system and enables adaptation to battery aging. It takes terminal voltage and current as inputs to estimate parameters, particularly the OCV. The battery dynamics are transformed into a linear format, which decouples the impact of current bias on OCV estimation. The specifics of the Parameter Estimation Module are elaborated in Section II.B.
2. **OCV-H-SOC Inversion Module (OIM):** Utilizing the OCV ($OCV_{Est}$) estimated by the Parameter Estimation Module, the OCV-H-SOC Inversion Module computes the hysteresis factor and determines the SOC ($SOC_{OCV-H}$) based on the intrinsic OCV-H-SOC map of LFP cells. A more detailed description of this module is provided in Section II.C.
3. **Condition Evaluation Module (CEM):** To ensure the reliability of SOC estimations, the Condition Evaluation Module adaptively quantifies the covariance ($cov_{SOC_{OCV-H}}$), which ascertains the confidence levels of $SOC_{OCV-H}$ using Fisher information and the LFP cell OCV-SOC relationship. The intricacies of the evaluation process are detailed in Section II.D.
4. **SOC Fusion Module (SFM):** The final stage of the estimation system is the SOC Fusion Module. This module takes $cov_{SOC_{OCV-H}}$ and $SOC_{OCV-H}$ as inputs and fuses them with the Coulomb counting method through a Kalman filter. This module smooths out potential noise and corrects for biases. The operational framework and benefits of the SOC Fusion Module are further discussed in Section II.E.

Table 1 The model parameter structure.

| Module | Inputs | Outputs | Parameters |
|---|---|---|---|
| PEM | $V_T, I$ | $OCV_{est}$ | $\lambda_0, \lambda_1, N$ |
| OIM | $OCV_{est}, I$ | $H, SOC_{OCV-H}$ | $C, f_{SOC}$ |
| CEM | $V_T, I, H, OCV_{est}, SOC_{est}$ | $Cov_{OCV-H}$ | $\sigma_{V_T}, f_{\frac{dSOC}{dOCV}}, N$ |
| SFM | $SOC_{OCV-H}, Cov_{OCV-H}, I$ | $SOC_{est}$ | $C_P, \Delta t, v_I$ |

## B. Parameter Estimation Module

The objective of the parameter estimation module is to estimate the battery parameters, particularly the OCV, in real time. To achieve this, we formulate a linear-in-the-parameters model, which facilitates the application of conventional parameter identification techniques, including the Kalman filter, recursive least squares, and linear regression [36].

As shown in Fig. 2, the battery dynamics can be presented as a second-order ECM with the following equations,

$$\dot{V}_1(t) = -\frac{1}{R_1 C_1} V_1(t) + \frac{I(t)}{C_1} \quad (1)$$

$$\dot{V}_2(t) = -\frac{1}{R_2 C_2} V_2(t) + \frac{I(t)}{C_2} \quad (2)$$

$$V_T(t) = OCV - R_0 I(t) - V_1(t) - V_2(t) \quad (3)$$

The ECM consists of an OCV, an ohmic resistor ($R_0$), and two Resistance-Capacitance (RC) pairs, denoted as ($R_1$ and $C_1$) and ($R_2$ and $C_2$). The current flowing through the ECM is represented by $I$ (in Amperes), while $V_1$ and $V_2$ indicate the voltages across the two RC pairs, respectively. The terminal voltage ($V_T$) serves as the output of the ECM.

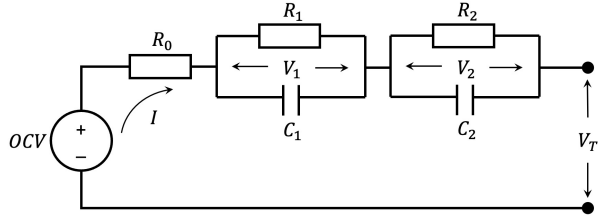

Fig. 2. Second-order ECM for battery.

By assuming that values of OCV, $R_0$, $C_1$ and $C_2$ remain constant during each parameter estimation step, we can apply the Laplace transform to the aforementioned equations, yielding the following equation after some substitution,

$$V_T(s) = OCV - \tau_1\tau_2 R_0 s^2 I(s) - (R_0\tau_1 + R_0\tau_2 + R_1\tau_2 + R_2\tau_1)sI(s) - (R_0 + R_1 + R_2)I(s) - \tau_1\tau_2 s^2 V_T(s) - (\tau_1 + \tau_2)sV_T(s) \quad (4)$$

where $s$ is the complex Laplace variable, and $\tau_1$ and $\tau_2$ represent the time constants of the two RC pairs, calculated as $R_1 C_1$ and $R_2 C_2$ respectively. We can then rearrange the equation into the following form,

$$V_T = [OCV \quad a \quad b \quad c \quad d \quad e] \begin{bmatrix} 1 \\ -\ddot{I} \\ -\dot{I} \\ -I \\ -\ddot{V}_T \\ -\dot{V}_T \end{bmatrix} \quad (5)$$

where the parameters $a$, $b$, $c$, $d$, and $e$ are defined as follows:

$$\begin{cases} a = \tau_1\tau_2 R_0 \\ b = R_0\tau_1 + R_0\tau_2 + R_1\tau_2 + R_2\tau_1 \\ c = R_0 + R_1 + R_2 \\ d = \tau_1\tau_2 \\ e = \tau_1 + \tau_2 \end{cases} \quad (6)$$

In this representation, the battery dynamics are transformed into a linear format. More detailed mathematical derivations are presented in Appendix V.A. By utilizing the measured terminal voltage $V_T$, its first and second derivatives $\dot{V}_T$ and $\ddot{V}_T$, along with the current $I$ and its derivatives $\dot{I}$ and $\ddot{I}$, we can estimate the parameters OCV, $a$, $b$, $c$, $d$, and $e$. Notably, in the linear format, the OCV is not directly related to the current ($I$). This implies that if there is a current bias, the term $c$, which represents the sum of the resistances of the battery, would absorb its impact. This property is critical, as it enables our SOC estimation scheme to be robust to current measurement bias.

In practical scenarios, the requirement for derivatives of measured signals presents a challenge due to measurement noise. To address this, we apply a second-order filter with stable poles (by setting $\lambda_0$, $\lambda_1 > 0$) to the measured signals. This filter design smooths out high-frequency noise while preserving the essential dynamic behavior of the LFP cell. Thus, the poles of the filter are chosen to be faster than the dynamics of the battery parameters. Typically, the time constants of the filter are an order of magnitude smaller than the time constants of battery (on the order of seconds to minutes) [37]. The second-order filter is defined as,

$$\Lambda(s) = \frac{\lambda_0}{s^2 + \lambda_1 s + \lambda_0} \quad (7)$$

Then, we have,

$$V_T \Lambda(s) = [OCV \quad a \quad b \quad c \quad d \quad e] \begin{bmatrix} 1 \\ -\ddot{I} \\ -\dot{I} \\ -I \\ -\ddot{V}_T \\ -\dot{V}_T \end{bmatrix} \Lambda(s) \quad (8)$$

$$V_T G_0(s) = [OCV \quad a \quad b \quad c \quad d \quad e] \begin{bmatrix} 1 \\ -I\, G_2(s) \\ -I\, G_1(s) \\ -I\, G_0(s) \\ -V_T\, G_2(s) \\ -V_T\, G_1(s) \end{bmatrix} \quad (9)$$

$$\begin{cases} G_0(s) = \dfrac{\lambda_0}{s^2 + \lambda_1 s + \lambda_0} \\ G_1(s) = \dfrac{\lambda_0 s}{s^2 + \lambda_1 s + \lambda_0} \\ G_2(s) = \dfrac{\lambda_0 s^2}{s^2 + \lambda_1 s + \lambda_0} \end{cases} \quad (10)$$

where $G_0(s)$, $G_1(s)$, and $G_2(s)$ are three transfer functions which can be further discretized using zero-order hold and applied to the measurement signals, $V_T$ and $I$. The discretized transfer functions can be represented as $G_0(z)$, $G_1(z)$, and $G_2(z)$ in the z-domain. The filtered and discretized measurement signals are then used for the paramter identification as follows,

$$\hat{V}_T[k] = [OCV \quad a \quad b \quad c \quad d \quad e] \begin{bmatrix} 1 \\ -\hat{I}''[k] \\ -\hat{I}'[k] \\ -\hat{I}[k] \\ -\hat{V}_T''[k] \\ -\hat{V}_T'[k] \end{bmatrix} \quad (11)$$

where $\hat{V}_T[k]$, $\hat{V}_T'[k]$, and $\hat{V}_T''[k]$ are the filtered and discretized terminal voltage and its first and second derivatives, and $\hat{I}[k]$, $\hat{I}'[k]$, and $\hat{I}''[k]$ are the filtered and discretized current and its first and second derivatives at the time step $k$. More detailed mathematical derivations are included in Appendix V.B.

Utilizing filtered and discretized signals as inputs, along with the derived linearized battery model, enables real-time estimation of battery OCV (denoted as $OCV_{est}$) using standard parameter identification techniques such as the Kalman Filter (KF) and Recursive Least Squares (RLS) with forgetting factors. In terms of battery dynamics, SOC reflects the long-term energy state of the system, changing more gradually than voltage or current, which can fluctuate rapidly. For most users, SOC estimation does not require millisecond- or even second-level updates, only an accurate, stable indication of available energy is needed. Furthermore, high-frequency updates using noisy signals may degrade accuracy by amplifying measurement noise.

Therefore, in this study, we adopt a batch least-squares estimation approach using a 100-second moving window. While recursive techniques like RLS and KF are fully compatible with our framework, we chose the moving window approach due to its practical simplicity and tighter integration with the overall system structure. First, it reduces implementation complexity, requiring only the window length as a tuning parameter, unlike RLS or KF, which need tuning of initial covariances, process and measurement noise models, and forgetting factors. Second, the moving window structure maintains consistency with the Condition Evaluation Module (Section II.D), which computes the Fisher Information Matrix over the same time window to evaluate parameter observability and SOC estimation confidence. Switching to RLS or KF would require fundamental redesign of this calculation, such as implementing weighted sensitivity matrices or incorporating time-varying decay.

The 100-second window size was selected after a sensitivity study balancing stability, responsiveness, and computational cost. Shorter windows were more responsive but prone to noise, while longer ones offered stability at the cost of delay. A 100-second horizon (100 samples at 1 Hz) offered the best trade-off across all tested scenarios, enabling accurate, stable, and robust parameter estimates.

*C. OCV-H-SOC Inversion Module*

As shown by the experimental data in Fig. 3, there a significant hysteresis between the charging and discharging OCV of a LFP battery cell. This section presents the module for estimating the SOC of LFP cells, incorporating the relationship between OCV, hysteresis, and SOC. This module is referred to as the "OCV-H-SOC Inversion Module" because it inverts the relationship between OCV, hysteresis, and SOC to determine SOC from given OCV and hysteresis information.

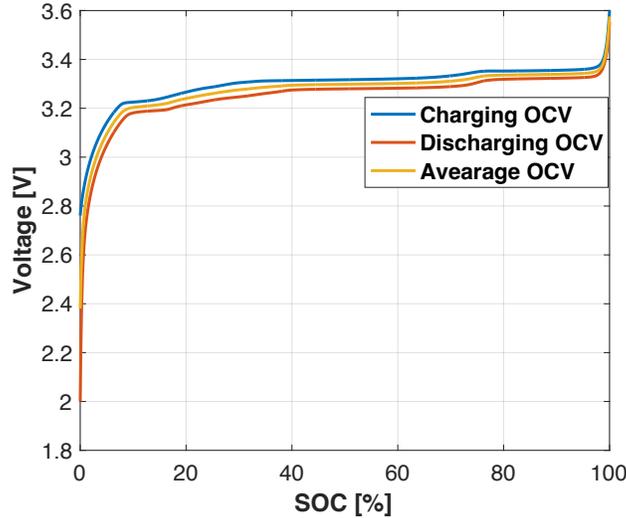

Fig. 3. LFP cell SOC-OCV curves were obtained using discharge and charging currents at a 1/50 C-rate and 25°C for a LithiumWerks APR18650M1-B cell (3.3V, 1.2 Ah LFP Battery).

The hysteresis observed in LFP cells during charging and discharging can be primarily attributed to the movement of phase boundaries between Li-rich ($LiFePO_4$) and Li-poor ($FePO_4$) phases [38], [39]. This phase transition behavior, driven by intercalation and deintercalation processes, causes asymmetrical voltage responses during charge and discharge cycles, leading to observable hysteresis in the OCV [40], [41]. This transition introduces energy barriers that result in a path-dependent voltage response, which must be captured for accurate SOC estimation.

To account for this, we adopt a recursive hysteresis model structure initially proposed by Plett [42]. This model have been extensively validated both analytically and experimentally in the literature [43], [44], [45]. However, to improve integration within our real-time SOC estimation framework, we simplify the model by removing the explicit hysteresis voltage term and instead introducing a hysteresis factor, $H$, that captures the memory and directionality of phase boundary motion. This hysteresis factor is dynamically updated as:

$$H(k) = \exp\left(-\left|\frac{I(k-1)}{C}\right|\right)H(k-1) + \left(1 - \exp\left(-\left|\frac{I(k-1)}{C}\right|\right)\right)sign(-I(k-1)) \qquad (12)$$

Here, $I(k-1)$ is the current at the previous time step, $C$ is a fitting parameter that determines how responsive the hysteresis state is to current magnitude, and the sign$(-I(k-1))$ term indicates the direction of the current (positive for discharging, negative for charging). The model uses an exponential term to weight the previous hysteresis state by a factor that decays with an increasing magnitude of current, thus providing a memory-like effect that captures the inertia in phase boundary movements. The sign function adjusts whether the hysteresis increases or decreases, depending on whether the cell is being charged or discharged. The model also captures how quickly the battery responds to changes in operational conditions, with a faster response at higher currents due to reduced exponential weighting. This formulation mimics the underlying phase boundary dynamics and allows the SOC-OCV-H mapping to reflect the true battery behavior across transitions.

The hysteresis factor ranges between -1 and 1, corresponding to fully discharging- and fully charging-dominated phases, respectively. This approach offers a computationally efficient and physically meaningful way to incorporate hysteresis into our estimation system. The associated phase transition dynamics between Li-rich and Li-poor regions that give rise to this behavior are illustrated in Fig. 4. In real-world BMS implementations, the hysteresis factor H is continuously updated even during parking conditions, as the BMS typically remains operational. The initial value of H is set using the final value from the previous cycle. If unavailable, the low-pass filter nature of the update equation ensures fast convergence once the system resumes operation.

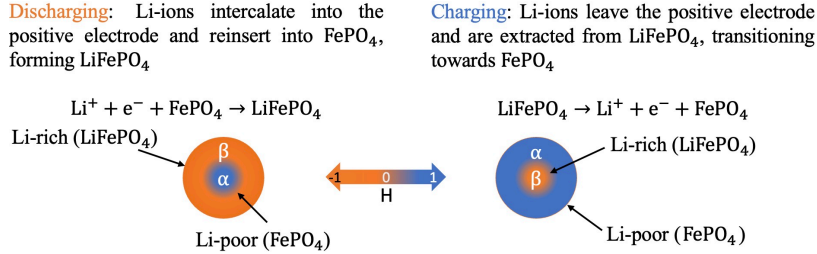

Fig. 4. Phase transition between Li-rich and Li-poor phases.

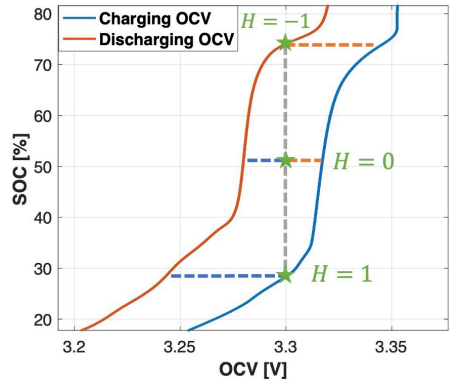

Fig. 5. Sample case of the SOC estimation based on OCV and H.

As shown in Fig. 5, determining SOC from an estimated OCV value depends on the hysteresis factor, $H$. For instance, with $H = 1$, indicating a charging-dominated case, an OCV of 3.3V corresponds to a SOC of 28.22%. Conversely, for $H = -1$, reflective of a discharging-dominated case, the same OCV of 3.3V corresponds to a SOC of 74.11%. When $H = 0$, suggesting an equilibrium state within the hysteresis loop, the SOC is 51% for an OCV of 3.3V. These relationships enable us to construct a 3D mapping between SOC, OCV, and the hysteresis factor using experimental data, as shown in Fig. 6.

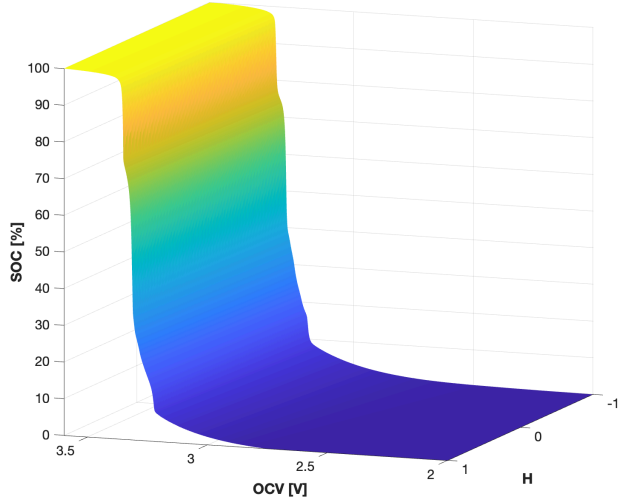

Fig. 6. OCV-H-SOC relationship 3D map, $f_{SOC}(OCV, H)$.

In the OCV-H-SOC inversion module, the hysteresis term $H$ is estimated in open-loop via Eq. (12), using the current as input. Then, utilizing the estimated OCV (denoted as $OCV_{est}$) from the parameter estimation module, along with the hysteresis term, we employ the SOC-OCV-H relationship 3D map to determine the SOC (denoted as $SOC_{OCV-H}$) as follows,

$$SOC_{OCV-H}(k) = f_{SOC}(OCV_{est}(k), H(k)) \tag{13}$$

*D. Condition Evaluation Module*

Like other parameter estimation methods mentioned in the literature review, the precision of our parameter estimation module is closely tied to the input current excitation levels. Specifically, the estimated OCV can become inaccurate under scenarios that lack input current excitation.

To address this challenge, we employ Fisher information and the Cramér-Rao bound to scrutinize the quality of the current excitation levels [46]. This evaluation enables us to quantify parameter uncertainty. Besides, when the estimation system operates in the SOC-OCV flat zone, roughly from 20% to 95% SOC, the SOC estimation uncertainty is heightened. Thus, this module quantities the covariance and confidence levels of the SOC estimated from the *OCV-H-SOC Inversion Module*.

According to Eq. (11), with the filtered and discretized terminal voltage as the measurement output, we determine the sensitivity, $S(k)$, of the voltage to each parameter by computing the partial derivatives of the output with respect to each parameter,

$$S(k) = \begin{bmatrix} \frac{\partial \widehat{V_T}}{\partial OCV}\bigg|_{k-N} & \frac{\partial \widehat{V_T}}{\partial a}\bigg|_{k-N} & \frac{\partial \widehat{V_T}}{\partial b}\bigg|_{k-N} & \frac{\partial \widehat{V_T}}{\partial c}\bigg|_{k-N} & \frac{\partial \widehat{V_T}}{\partial d}\bigg|_{k-N} & \frac{\partial \widehat{V_T}}{\partial e}\bigg|_{k-N} \\ \frac{\partial \widehat{V_T}}{\partial OCV}\bigg|_{k-N+1} & \frac{\partial \widehat{V_T}}{\partial a}\bigg|_{k-N+1} & \frac{\partial \widehat{V_T}}{\partial b}\bigg|_{k-N+1} & \frac{\partial \widehat{V_T}}{\partial c}\bigg|_{k-N+1} & \frac{\partial \widehat{V_T}}{\partial d}\bigg|_{k-N+1} & \frac{\partial \widehat{V_T}}{\partial e}\bigg|_{k-N+1} \\ \vdots & \vdots & \vdots & \vdots & \vdots & \vdots \\ \frac{\partial \widehat{V_T}}{\partial OCV}\bigg|_{k} & \frac{\partial \widehat{V_T}}{\partial a}\bigg|_{k} & \frac{\partial \widehat{V_T}}{\partial b}\bigg|_{k} & \frac{\partial \widehat{V_T}}{\partial c}\bigg|_{k} & \frac{\partial \widehat{V_T}}{\partial d}\bigg|_{k} & \frac{\partial \widehat{V_T}}{\partial e}\bigg|_{k} \end{bmatrix} \tag{14}$$

Then, we have,

$$S(k) = \begin{bmatrix} 1 & -\hat{I}''[k-N] & -\hat{I}'[k-N] & -\hat{I}[k-N] & -\widehat{V_T}''[k-N] & -\widehat{V_T}'[k-N] \\ 1 & -\hat{I}''[k-N+1] & -\hat{I}'[k-N+1] & -\hat{I}[k-N+1] & -\widehat{V_T}''[k-N+1] & -\widehat{V_T}'[k-N+1] \\ \vdots & \vdots & \vdots & \vdots & \vdots & \vdots \\ 1 & -\hat{I}''[k] & -\hat{I}'[k] & -\hat{I}[k] & -\widehat{V_T}''[k] & -\widehat{V_T}'[k] \end{bmatrix} \tag{15}$$

where $\hat{V}_T[k]$, $\hat{V}_T'[k]$, and $\hat{V}_T''[k]$ are the filtered and discretized terminal voltage and its first and second derivatives, and $\hat{I}[k]$, $\hat{I}'[k]$, and $\hat{I}''[k]$ are the filtered and discretized current and its first and second derivatives at the time step $k$, as mentioned in Section II.C. $N$ is the number of data points/time steps used in the linear regression of the online parameter estimation. The Fisher information matrix, $F(k)$, is,

$$F(k) = \frac{S(k)^T S(k)}{\sigma_{V_T}^2} \tag{16}$$

Here, $\sigma_{V_T}$ is the terminal voltage measurement noise covariance.

While the F(k) is ideally invertible, it may become ill-conditioned or singular under low excitation conditions. A large moving window can help alleviate this by including more informative data points. Besides, in practice, Tikhonov regularization [47] is applied to ensure numerical stability:

$$F(k) = F(k) + \epsilon I_{d \times d} \tag{17}$$

Where $\epsilon = 10^{-8}$ is a very small constant, and $I_{d \times d}$ is the d × d identity matrix matching the size F(k). This regularized matrix guarantees invertibility.

The Cramér-Rao bound provides the lower bound of the parameter estimation covariance, computed from the inverse of the Fisher information matrix [48]. Hence, we determine the lower bound of the estimated OCV's covariance by selecting the first element in the inverse Fisher information matrix,

$$Cov_{OCV} \geq [F(k)^{-1}]_{11} \tag{18}$$

Considering the nearly monotonic OCV-SOC relationship depicted in Fig. *6*, and assuming no uncertainty related to the hysteresis factor, we can approximate the covariance of the estimated SOC from OCV as,

$$Cov_{SOC} = \left(\frac{dSOC}{dOCV}\right)^2 Cov_{OCV} \tag{19}$$

where $\frac{dSOC}{dOCV}$ is the inverse slope of the SOC-OCV curve, and is a function of H and SOC,

$$\frac{dSOC}{dOCV} = f_{\frac{dSOC}{dOCV}}(H, SOC) \tag{20}$$

This derivation is a critically important attribute of the algorithm. Namely, in flat zones of the OCV-SOC curve, uncertainty in OCV produces amplified uncertainty in the estimated SOC from the inversion process. As illustrated in Fig. 7, we can see that the value of $\frac{dSOC}{dOCV}$ is extremely high when H is close to 1 and SOC is around 85%. This is because when charging is dominant, the SOC-OCV curve around 85% SOC is almost purely flat ($\frac{dOCV}{dSOC} \approx 0$). Namely, the SOC ($SOC_{OCV-H}$) obtained from the OCV-H-SOC Inversion Module is not reliable in this case.

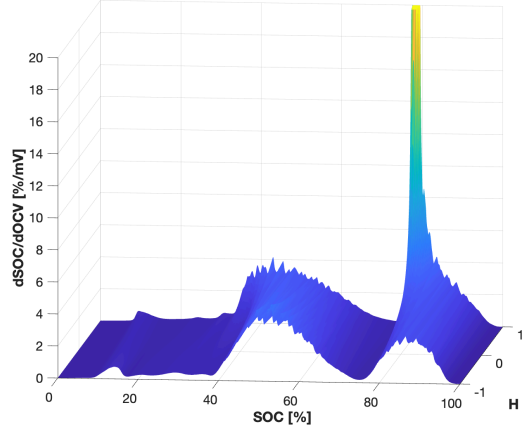

Fig. 7. The inversion of the slope of the SOC-OCV curve with the given estimated H and SOC.

According to above equations, we can then estimate the covariance and confidence levels of $SOC_{OCV-H}$ with the following relation,

$$Cov_{SOC_{OCV-H}}(k) \geq \left( f_{\frac{dSOC}{dOCV}}(H(k), SOC_{est}(k-1)) \right)^2 [F(k)^{-1}]_{11} \tag{21}$$

where $Cov_{SOC_{OCV-H}}$ is the covariance of estimated $SOC_{OCV-H}$, and $SOC_{est}(k-1)$ is the estimated SOC of the system from the SOC Fusion Module at the previous time step. Since even at a 3C charging current, the change in SOC within one second (the time step used in this study) is less than 0.1%, we assume that $SOC_{est}(k-1)$ is sufficiently close to the true SOC to accurately determine $f_{\frac{dSOC}{dOCV}}$.

Based on Fig. 7 and Eq. (19), we expect that the covariance of estimated SOC from OCV-H-SOC Inversion Module ($Cov_{SOC_{OCV-H}}$) is high when the battery is in the SOC-OCV flat zone, where the value of $f_{\frac{dSOC}{dOCV}}$ is substantially high. Additionally, when the current excitation level is low, then the first element of the Fisher information matrix $[F(k)^{-1}]_{11}$ is high, which increases the covariance. In summary, the Condition Evaluation Module assesses the estimated SOC covariance from OCV-H-SOC Inversion Module, under varying operational conditions.

*E. SOC Fusion Module*

The SOC Fusion Module serves as the integrative core of our system. It combines the traditional Coulomb counting with the estimated SOC from the *OCV-H-SOC Inversion Module*, via a Kalman filter framework. The module enhances SOC accuracy by performing the following update cycle:

**Model Prediction:**
Coulomb counting is utilized for real-time SOC prediction by integrating current flow over time:

$$SOC_{cc}(k) = SOC_{est}(k-1) + \frac{I(k-1)\Delta t}{C_P} \tag{22}$$

where $SOC_{cc}(k)$ is the SOC estimated by Coulomb counting, $I(k-1)$ is the current at the previous step, $\Delta t$ is the sampling time, and $C_P$ is the battery capacity, which is assumed to be known in advance.

**Measurement update:**
The SOC estimation from the OCV and hysteresis model, $SOC_{OCV-H}(k)$, is used as the measurement part of the Kalman filter.

**Error Covariance Prediction**:

$$P_p(k) = P_m(k-1) + v_I \tag{23}$$

The predicted error covariance, $P_p(k)$, accounts for process noise, and $v_I$ captures uncertainties in the model prediction phase. It corresponds to the uncertainty in current measurement.

**Kalman Gain Calculation**:
$$K(k) = \frac{P_p(k)}{P_p(k) + cov_{SOC_{OCV-H}}(k)} \tag{24}$$

The Kalman gain, $K(k)$, determines the weighting of the measurement update relative to the predicted state, influenced by the confidence in the $SOC_{OCV-H}$ estimates provided by the Condition Evaluation Module.

**State Update**:
$$SOC_{est}(k) = SOC_{cc}(k) + K(k)\big(SOC_{OCV-H}(k) - SOC_{cc}(k)\big) \tag{25}$$

The state update refines the SOC estimate, $SOC_{est}(k)$, by reconciling the differences between the Coulomb counting (model prediction) and the $SOC_{OCV-H}(k)$ (measurement correction).

**Covariance Update**:
$$P_m(k) = \big(I - K(k)\big)P_p(k) \tag{26}$$

The updated measurement error covariance, $P_m(k)$, reflects the reduction in uncertainty following the measurement update.

While Coulomb counting offers a straightforward estimation technique by cumulative current integration, it is prone to drift from initial SOC inaccuracies and current measurement biases. Within the Kalman filter, however, Coulomb counting serves as the model, providing a baseline SOC trajectory. It is updated using $SOC_{OCV-H}$, which, despite being noisy, corrects Coulomb counting via OCV-H inversion.

The process noise of the Kalman filter is interpreted as the uncertainty of the current measurement. Meanwhile, the measurement noise, indicated by the covariance of $SOC_{OCV-H}$, modulates the confidence in updates according to the latest condition assessments. By blending Coulomb counting with $SOC_{OCV-H}$ estimations, the SOC Fusion Module produces a synchronized SOC output, $SOC_{est}(k)$. Such calibration with real-time data allows the system to deliver a smooth SOC estimation, effectively correcting for current biases and initial SOC inaccuracies. In addition, because the Kalman filter does not rely on other battery parameters, it is less impacted by the aging and temperature effects compared to other methods mentioned in the literature review section.

Ultimately, the proposed algorithm can be viewed as a direct evolution of the SOC estimation algorithm in the 2004 paper by Verbrugge and Tate at General Motors [49]. This work similarly fused Coulomb counting and OCV inversion, while accounting for hysteresis, but for nickel metal-hydride (NiMH) batteries.

### III. MODEL PERFORMANCE

In this section, we begin by outlining the configuration of our test environment. We introduce a state-of-art benchmark method for comparison: the UKF approach. The proposed algorithm is then rigorously tested against the benchmark using driving cycle data and through a series of challenging scenarios. These scenarios mimic real-world complexities, including high initial SOC error, prolonged operation in SOC-OCV flat zones, current bias, voltage quantization error, low temperature, and insufficient current excitation. The results from these tests are discussed to provide a nuanced perspective on the robustness and accuracy of the proposed algorithm under various conditions.

*A. Testing Setup*

In this study, we utilize a LithiumWerks APR18650M1-B, a 3.3 V, 1.2 Ah LiFePO4 battery, for all tests. The ambient temperature around the battery cell is maintained at either 25°C or 10°C within an ESPEC BTL-433 environmental chamber. The battery is tested using an Arbin high-current cylindrical cell holder. Charge and discharge cycles, mimicking vehicle operations, are executed using a PEC SBT2050 cycler.

*B. Benchmark Approaches*

To validate the efficacy of our proposed SOC estimation system, we compare it against a well-established method, the UKF approach. The UKF represents an enhancement of the traditional Kalman filter, adept at capturing the non-linearities in system dynamics without the need for linearization. In this study, parameters of the battery's 2RC model for the UKF are determined using the Hybrid Pulse Power Characterization (HPPC) test, which involves 1C discharge pulses at 5% SOC intervals. All tests are conducted at a stable ambient temperature of 25°C across various SOC levels. The SOC-OCV curve utilized in the UKF is derived from an average of the charging and discharging SOC-OCV curves, specifically obtained at 25°C. The benchmark method is tested under identical testing conditions to those used for our proposed algorithm to ensure a fair and consistent comparative analysis.

In addition to the UKF, we include a comprehensive benchmarking study presented in Appendix V.C, where we implement and evaluate two leading data-driven models, LSTM and decoder-only Transformer model, both as standalone predictors and in combination with Coulomb Counting via a Kalman Filter. Appendix V.C details the model architectures, fusion strategies, test conditions, and computational costs. It also provides a comparative analysis across various realistic operating scenarios, highlighting the strengths and limitations of each method.

*C. Under Urban Driving Conditions*

For the ideal condition test, the battery is cycled from 100% SOC to 0% SOC and then charged back to 100% SOC using the Urban Dynamometer Driving Schedule (UDDS) driving cycle profile. We refer to these conditions as 'ideal' because the test is performed with no measurement errors, at a constant room temperature of 25°C, and under a highly dynamic current profile where persistency of excitation conditions is satisfied.

As illustrated in Fig. 8, the blue, red, and yellow lines represent the SOC from the UKF, proposed method, and true values, respectively. For the ideal condition test, both methods work very well. The Root Mean Square Error (RMSE) results for the proposed method and the UKF are 0.49% and 3.92%, respectively. For both methods, the initial SOC guess was set at 50% to introduce a 50% initialization error. The plot shows that both SOC estimation methods quickly converge to the true SOC value. This rapid convergence is primarily due to the true initial SOC value being 100%, which lies in the non-flat zone of the OCV-SOC curve. Additionally, a deliberately designed high initial SOC error covariance helps accelerate convergence in the presence of high initial uncertainty. However, although the UKF provides relatively good estimation results, its performance does not match that of the proposed method, which estimates battery parameters in real-time and accounts for the hysteresis phenomenon.

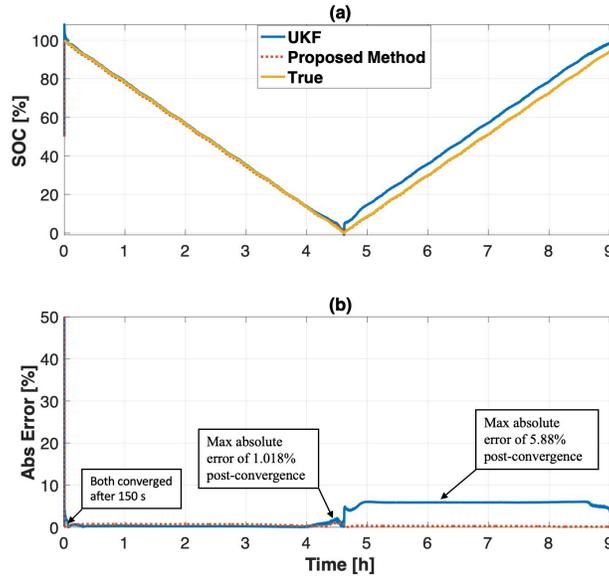

Fig. 8. Test results under ideal condition: No voltage and current measurement error, operation from 100% to 0% SOC at 25°C. (a) SOC comparison between the UKF, proposed method, and true SOC. (b) Absolute error (%) comparison, showing a maximum absolute error of 1.018% for the proposed method and 5.88% for the UKF after convergence. The RMSE values are 0.49% and 3.92% for the proposed method and UKF, respectively.

To better illustrate the operation of the proposed method, Fig. 9 presents the SOC estimated from OCV and hysteresis $SOC_{OCV-H}$ and its covariance $cov_{SOC_{OCV-H}}$ utilized in the SOC Fusion Module. In the subplot, it is noticeable that $SOC_{OCV-H}$ is noisy and inaccurate, particularly between 40.5% to 63.75% SOC and 78.2% to 93.15% SOC. This inaccuracy arises because these are super flat zones in the SOC-OCV curve. As illustrated in Fig. 7, the $\frac{dSOC}{dOCV}$ values are up to 20 %/mV when SOC is around 85% during charging, indicating that the SOC is insufficiently sensitive to variations in OCV values. Consequently, the proposed Condition Evaluation Module output notably high $cov_{SOC_{OCV-H}}$ values for the system in the super flat zones, as shown in the lower subplot of Fig. 9. Conversely, when the SOC is between 0% to 8.75% and 97.6% to 100%, the SOC-OCV curve exhibits its steepest slopes. During these intervals, the Condition Evaluation Module assigned low $cov_{SOC_{OCV-H}}$ values to the SOC Fusion Module. Thus, in this test case, the SOC Fusion Module relies more on $SOC_{OCV-H}$ when the slope of the SOC-OCV curve is steep and leans more towards Coulomb counting when the battery operates within the SOC-OCV flat zones, ensuring the fused SOC result ($SOC_{est}$) greatly matches the true SOC. It performs this weighting automatically, based on the mathematical model structure and data.

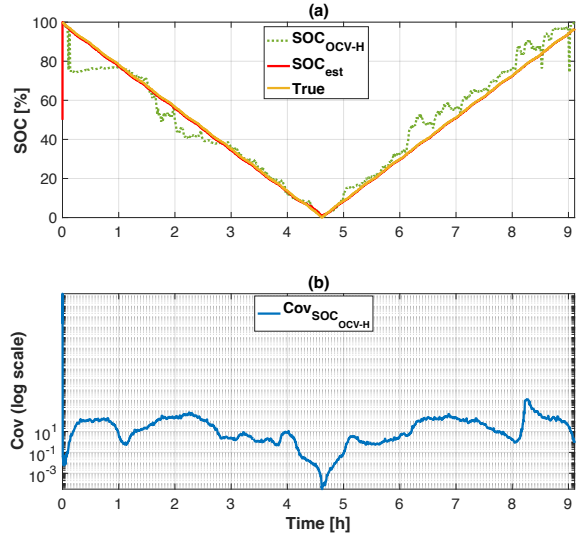

Fig. 9. Test result of the proposed method with the ideal condition. (a) SOC comparison between $SOC_{OCV-H}$ (*SOC estimated from OCV and hysteresis model*), $SOC_{est}$, and true SOC. (b) Covariance of SOC from the OCV-H-SOC inversion module on a logarithmic scale.

*D. Prolonged Operation in SOC-OCV Flat Zones*

In this subsection, we evaluate the performance of the proposed method and the UKF specifically within the SOC-OCV flat zones, for an extended period. The battery is charged from 20% SOC to 80% SOC and then discharged back to 20% SOC. Four different driving cycle profiles, including the Orange County Transit Bus Cycle (OCTBC), the California Unified Cycle (OCTBC), the US06 Drive Cycle, and the New York City Cycle (NYCC), are used to generate the current profiles for validating system performance under different operational conditions. The corresponding voltage and current profiles are shown in Fig. 10.

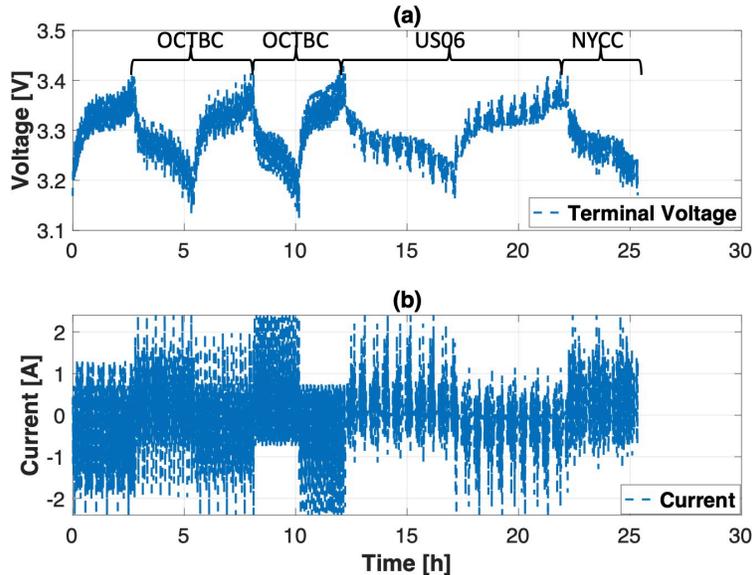

Fig. 10. Voltage and Current profiles using the Orange County Transit Bus Cycle, the California Unified Cycle, the US06 Drive Cycle, and the New York City Cycle. (a) Voltage profile. (b) Current profile.

The test spans approximately 24 hours and is exclusively conducted within the flat zone range from 20% SOC to 80% SOC to thoroughly assess the impact of SOC-OCV flat zones on both the proposed method and the UKF. The true initial SOC is set at 20%, while an erroneous initial guess of 100% SOC is intentionally set to introduce an 80% initial error.

The comparison results are illustrated in Fig. 11. The RMSE for the proposed method and the UKF are 2.54% and 6.69%, respectively. The RMSE values for both methods are higher than those observed in ideal conditions, because the flat SOC-OCV zones reduce opportunities for correction that are available in steeper zones. Despite this, the proposed method still performs better than the UKF.

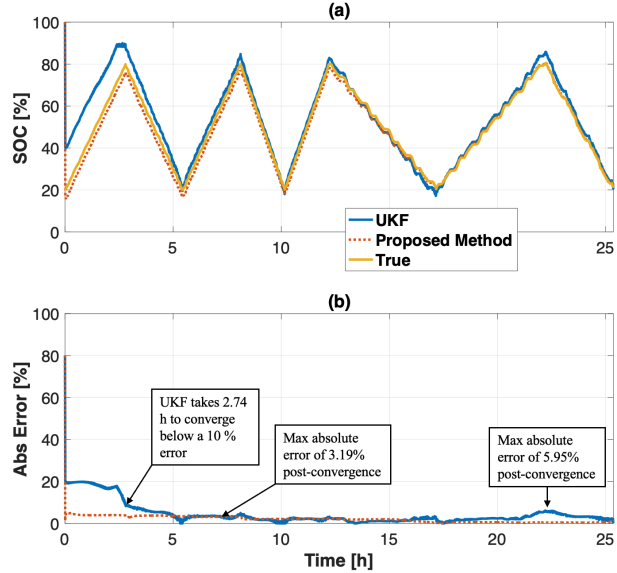

Fig. 11. Test results of the SOC-OCV flat zones (operation from 20% to 80% SOC at 25°C with four different current profiles). (a) SOC comparison between the UKF, proposed method, and true SOC. (b) Absolute error (%) comparison, showing a maximum absolute error of 3.19% for the proposed method and 5.95% for the UKF after convergence. The RMSE values are 2.54% and 6.69% for the proposed method and UKF, respectively.

The state-space equations used by the UKF are:

$$\begin{bmatrix} SOC(k+1) \\ V_1(k+1) \\ V_2(k+1) \end{bmatrix} = \begin{bmatrix} 1 & 0 & 0 \\ 0 & e^{-\frac{\Delta t}{R_1 C_1}} & 0 \\ 0 & 0 & e^{-\frac{\Delta t}{R_2 C_2}} \end{bmatrix} \begin{bmatrix} SOC(k) \\ V_1(k) \\ V_2(k) \end{bmatrix} + \begin{bmatrix} -\frac{\Delta t}{C_P} \\ \left(1 - e^{-\frac{\Delta t}{R_1 C_1}}\right) R_1 \\ \left(1 - e^{-\frac{\Delta t}{R_2 C_2}}\right) R_2 \end{bmatrix} I(k) + w_{UKF} \quad (27)$$

$$V_T(k) = OCV(SOC) - I(k)R_0 - V_1(k) - V_2(k) + v_{UKF} \quad (28)$$

where $R_1$, $R_2$, $C_1$, and $C_2$ are the fitted battery RC parameters (which change crossing SOC), $V_1$ and $V_2$ are the voltages across the two RC pairs, $C_P$ is the battery capacity, and $\Delta t$ is the sampling time. $V_T$ is the terminal voltage and $R_0$ is the ohmic resistance. Besides, $w_{UKF}$ and $v_{UKF}$ are process and measurement noises that capture model inaccuracies and measurement errors, respectively.

For the UKF, convergence to below a 10% SOC estimation error takes 2.74 hours because of its slower adjustment rate. In the flat OCV zones, reliance on predefined, nonlinear parameters makes the UKF less sensitive to OCV variations. Changes in OCV may be masked by inaccurate RC parameters or absorbed through adjustments in $V_1$ and $V_2$ with the noise terms $w_{UKF}$ and $v_{UKF}$. In contrast, the proposed system is more sensitive to OCV changes. It transforms battery dynamics into a linear format, making it more responsive to OCV variations. It also updates all parameters in real time, ensuring the model quickly adapts and stays aligned with the actual battery behavior.

The corrective effort of the UKF is primarily represented by the Kalman gain for SOC and the voltage difference between the true and predicted terminal voltages. As depicted in Fig. 10, for the UKF, the voltage error, along with the Kalman gain for SOC, drops significantly after a few iterations. The high initial SOC covariance results in a high Kalman gain at the beginning, which, combined with the high initial voltage error, leads to fast corrections initially. However, when the gain is low and the voltage difference is small due to the flat zone effect, the gain applied to the output error injection is small. Thus, the UKF essentially resembles a Coulomb counting method and converges very slowly. It is also notable that a higher process noise setup for the Coulomb counting part can help with the convergence speed. However, it also degrades the overall performance of the UKF because, within super flat zones, it must rely on Coulomb counting.

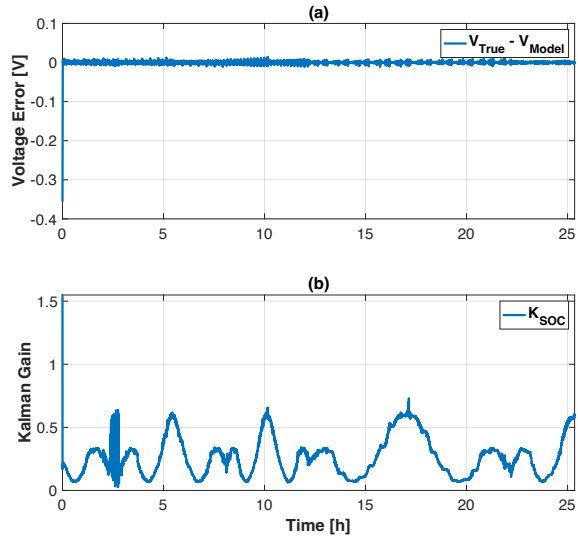

Fig. 12. Voltage prediction error and Kalman gain for the UKF method in the SOC-OCV flat zone case. (a) Error between measured and predicted terminal voltage in the UKF model. (b) Kalman gain for SOC correction in the UKF.

Conversely, as shown in Fig. 11, the SOC estimates from the proposed method converge quickly to the true SOC values, assisted by the SOC estimated from OCV-H inversion. Although the $SOC_{OCV-H}$ values exhibit considerable noise, the SOC Fusion Module, combined with $cov_{SOC_{OCV-H}}$ from the Condition Evaluation Module, effectively mitigates the impact of the very flat SOC-OCV zone and low current excitation levels, ensuring accurate and smooth estimation results. The detailed comparison between $SOC_{est}$, $SOC_{OCV-H}$, and the true SOC is presented in the upper subplot of Fig. 11. We can see that $SOC_{est}$ rapidly converges to the true SOC value after compensating for the initially high guess error. By cross-referencing the lower subplot of Fig. 13, the values of $SOC_{OCV-H}$ align with the true SOC when $cov_{SOC_{OCV-H}}$ is low. Consequently, the SOC Fusion Module effectively adjusts $SOC_{est}$ based on $SOC_{OCV-H}$ and $cov_{SOC_{OCV-H}}$, tailored to the specific operational conditions.

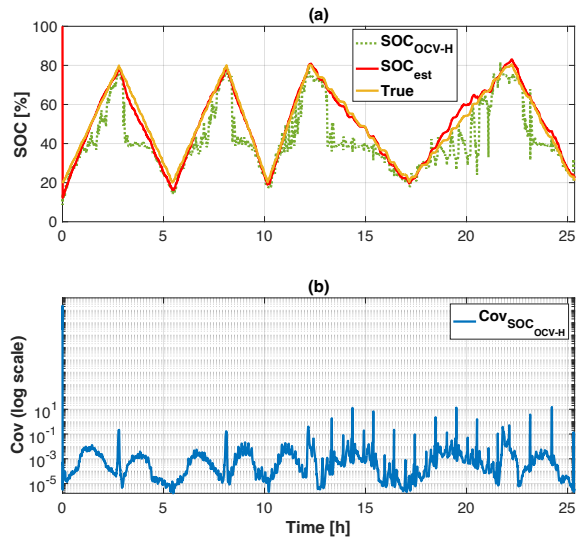

Fig. 13. Proposed method under the flat zone case. (a) SOC comparison between $SOC_{OCV-H}$ (SOC estimated from OCV and hysteresis model), $SOC_{est}$, and true SOC. (b) Covariance of SOC from the OCV-H-SOC inversion module on a logarithmic scale.

*E. Prolong Operation in SOC-OCV Flat Zones and with Current Bias*

Current bias is always a concern for battery state estimation, yet almost always ignored in the literature. In this subsection, the performance of the proposed method and UKF is evaluated under another extreme condition [50]. In addition to prolonged operation within the flat SOC-OCV zones with a high initial guess error, we apply a significant current bias of -0.05 A (the negative current indicates that charging here). This bias is substantial given the battery's capacity of only 1.2 Ah.

As illustrated in Fig. 12, the RMSE for the proposed method and the UKF are 2.99% and 15.44%, respectively. The proposed method significantly outperforms the UKF. The UKF, being a model-based method, relies on current measurements to update the SOC, voltages across two RC pairs, and terminal voltage. Consequently, the current measurement bias adversely affects the internal state estimates and terminal voltage prediction. Suppose we set a higher process noise for the Coulomb counting part to encourage the UKF to rely more heavily on the measurement correction. This sacrifices the performance of the method under other conditions, namely under high voltage noise, and the estimation accuracy would not significantly improve. For instance, the negative current bias increases the SOC calculated by Coulomb counting, leading to higher estimated terminal voltages and, subsequently, an overestimation of SOC. Although the UKF's internal model attempts to capture battery voltage dynamics to offset the cumulative error from Coulomb counting, the impact of current bias on internal voltages results in a significant positive SOC estimation offset of up to 26.39%.

On the other hand, the current bias has only a limited impact on the proposed method. Thanks to Eq. (8) within the Parameter Estimation Module, the current bias predominantly affects the parameter 'c', which is related to the internal resistances of the battery. The impact of current bias is absorbed by the estimation of internal resistances rather than affecting the estimated OCV. With fewer components impacted by the current bias, the proposed method maintains its performance by utilizing relatively accurately estimated OCV to mitigate accumulated SOC errors associated with Coulomb counting.

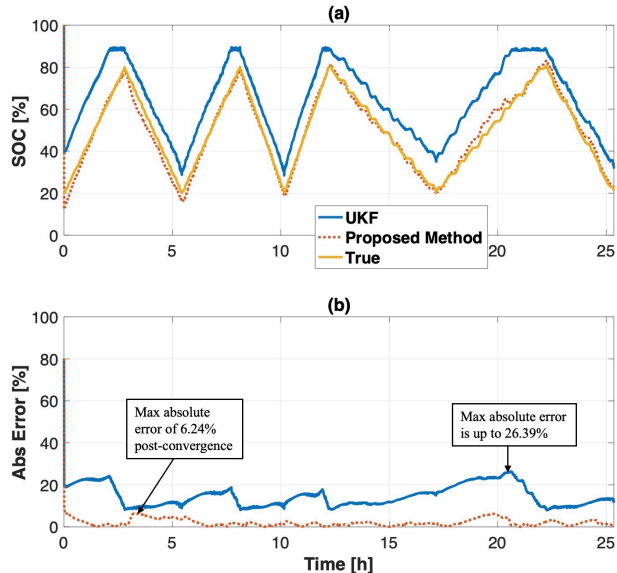

Fig. 14. Test results of the SOC-OCV flat zone adding a high current measurement bias. The operation is from 20% to 80% SOC at 25°C with four different current profiles and -0.05A current bias. (a) SOC comparison between the UKF, proposed method, and true SOC. (b) Absolute error (%) comparison, showing a maximum absolute error of 6.24% for the proposed method and 26.39% for the UKF after convergence. The RMSE values are 2.99% and 15.44% for the proposed method and UKF, respectively.

*F. Prolonged Operation in SOC-OCV Flat Zones with Voltage Quantization Errors*

Since the SOC-OCV curve of LFP cells is quite flat, there are concerns about the impacts of voltage measurement bias and sensor noise. Voltage measurement bias refers to a systematic offset in the measured voltage, meaning that the recorded voltage values consistently deviate from the true values by a fixed amount (positive or negative). In practice, voltage measurement sensors should be periodically calibrated and updated, thus the voltage measurement bias is not modeled here. Instead, we consider voltage sensor noise, i.e., ADC (analog to digital converter) Quantization Noise as a major error source. In this subsection, voltage quantization errors were introduced to the voltage measurement results to simulate the effects of a 10-bit analog-to-digital converter (ADC) [51] with a maximum 5V supply voltage for testing both methods.

The quantized terminal voltage, $V_Q$, is calculated by,

$$V_Q = \left\lfloor \frac{V_T}{\Delta V_Q} + 0.5 \right\rfloor \times \Delta V_Q \tag{29}$$

where $V_T$ is the measured terminal voltage and $\Delta V_Q$ is the quantization step size, defined as,

$$\Delta V_Q = \frac{V_{max}}{2^n - 1} \tag{30}$$

Here, $V_{max}$ represents the maximum voltage (5V) that the ADC can measure and $n$ is the resolution of the ADC, which determines the number of distinct levels (1024 for a 10-bit ADC). The impact of quantization error introduced by the ADC is depicted in Fig. 15, ranging from -2.44 mV to 2.44 mV. This variability in error magnitude correlates with the dynamics of the terminal voltage.

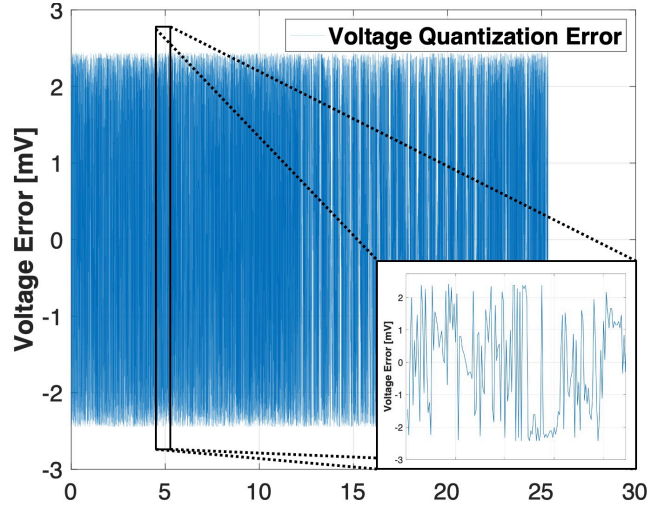

Fig. 15. Voltage quantization error added in the test.

The test results are displayed in Fig. 16. We observe that the voltage quantization error does not significantly impact either method. The RMSE for the proposed method and the UKF are 2.69% and 7.1325%, respectively. These errors are only slightly higher than those obtained without the presence of voltage quantization errors. This minimal impact is attributed to the distribution of the quantization error due to the dynamic nature of the voltage changes. As illustrated in Fig. 17, the voltage quantization errors are almost uniformly distributed with a zero-mean value. Since there is no constant voltage bias introduced by the ADC, these uniformly distributed measurement errors are effectively managed by both methods.

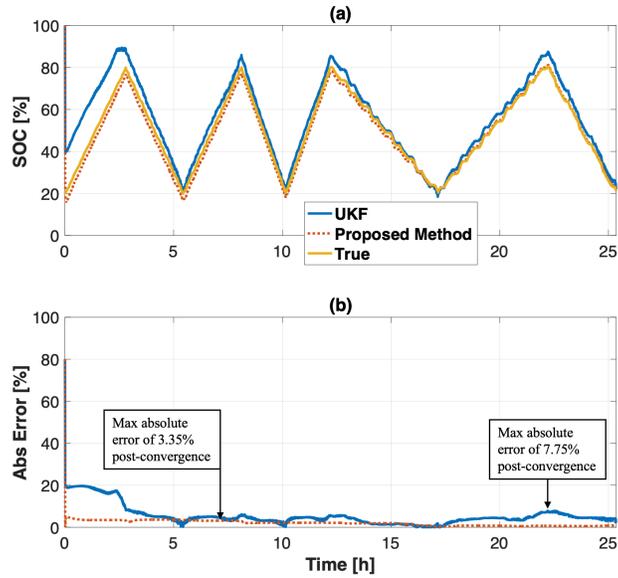

Fig. 16. Test results of the SOC-OCV flat zone with voltage quantization errors. The operation is from 20% to 80% SOC at 25°C with four different current profiles and 10-bit ADC with a maximum 5V supply voltage. (a) SOC comparison between the UKF, proposed method, and true SOC. (b) Absolute error (%) comparison, showing a maximum absolute error of 3.35% for the proposed method and 7.75% for the UKF after convergence. The RMSE values are 2.69% and 7.1325% for the proposed method and UKF, respectively.

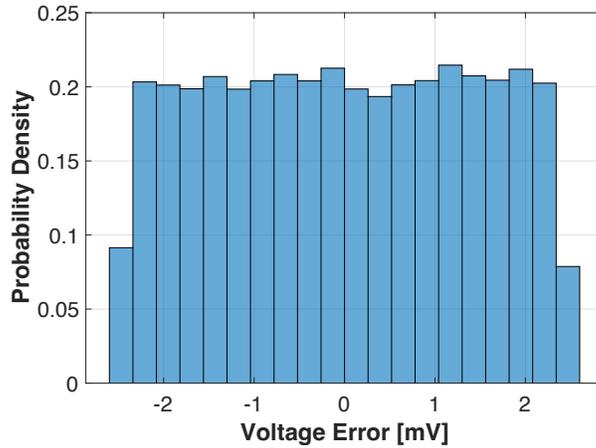

Fig. 17. Distribution of the voltage quantization error added in the test.

*G. Prolong Operation in SOC-OCV Flat Zones with Low Temperature*

In this subsection, we evaluate the performance of the proposed method and the UKF within the SOC-OCV flat zone at 10°C. The parameters for both methods are set based on data obtained at 25°C. Specifically, the OCV-SOC curves used for both methods and the RC parameters utilized in the UKF are derived from tests conducted at 25°C. This setup aims to assess how well each method performs under temperatures for which they do not have directly associated parameters.

The test results are presented in Fig. 18. The RMSE for the proposed method and the UKF are 3.28% and 28.11%, respectively. The UKF exhibits a peak error of up to 65%, primarily because its performance heavily depends on the accuracy of the RC parameters. When using RC parameters calibrated at 25°C, significant discrepancies arise at 10°C, akin to substantial offset errors in the model. The UKF struggles to manage these discrepancies as Gaussian-distributed process noise. Conversely, the proposed method is less impacted by the temperature change because it primarily relies on the OCV-SOC relationship, which is less sensitive to temperature variations compared to the RC parameters. Moreover, parameters are estimated in real-time via the *Parameter Estimation Module* in Section II.B.

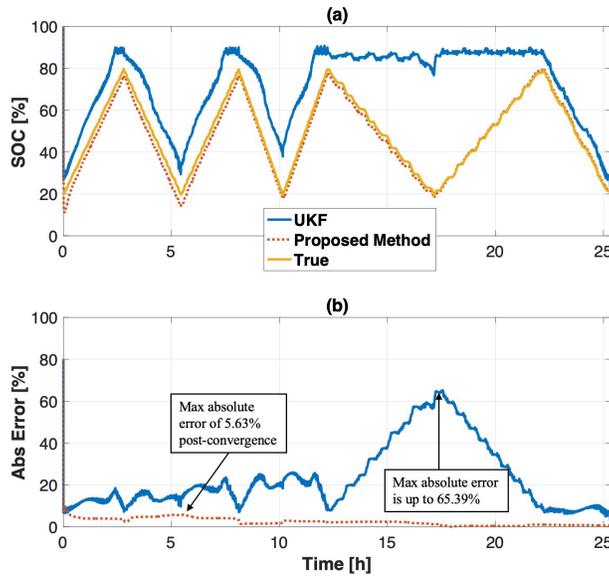

Fig. 18. Test results of the SOC-OCV flat zone and low temperature. The operation is from 20% to 80% SOC at 10°C with four different current profiles. (a) SOC comparison between the UKF, proposed method, and true SOC. (b) Absolute error (%) comparison, showing a maximum absolute error of 5.63% for the proposed method and 65.39% for the UKF after convergence. The RMSE values are 3.28% and 28.11% for the proposed method and UKF, respectively.

As depicted in the upper plot of Fig. 19, the LFP cell's OCV-SOC curves are very similar across different temperatures. This similarity might lead readers to question why the proposed method does not perform as well at 10°C as it does at 25°C. The answer is intricately linked to the flat zone of the OCV-SOC curve for the LFP cell. For a given OCV value, we calculated the SOC differences between 10°C and 25°C from both the discharge and charge curves. As illustrated in the lower plot of Fig. 19, within the flat zone, these SOC differences at a given OCV are still substantial, posing a challenge for accurately determining the SOC at 10°C using the 25°C OCV-SOC relationship. However, this issue can be effectively addressed in practical applications by incorporating temperature as an input factor and using a temperature-adaptive OCV-H-SOC relationship.

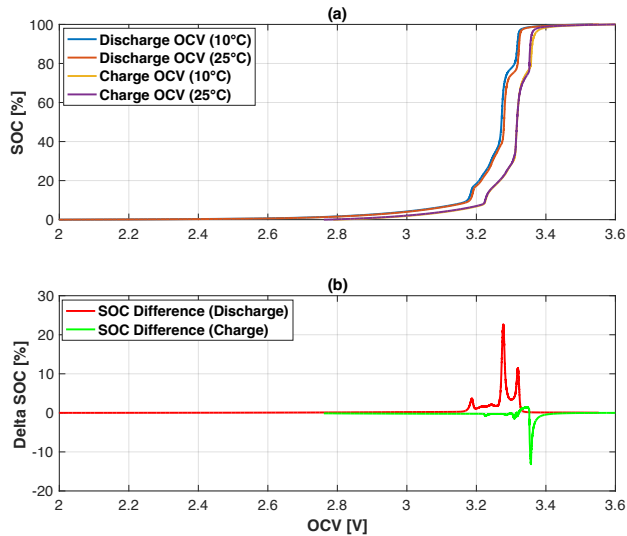

Fig. 19. OCV-SOC relationship of an LFP cell at 10°C and 25°C. (a) Discharge and charge OCV curves at 10°C and 25°C. (b) SOC differences due to temperature variations: the red line shows SOC differences between 10°C and 25°C at given discharging OCV values, while the green line shows SOC differences between 10°C and 25°C at given charging OCV values.

*H. Operation in SOC-OCV Flat Zones with a Long Constant Current Segment*
The last test aims to verify the ability of the proposed system to handle conditions with insufficient current excitation. Unlike online parameter estimation methods, the UKF, as a model-based method with predefined parameters, does not have these limitations.

Therefore, it is not included in this discussion. In the test, the initial SOC guess was set at 100%, while the true SOC was 80%, introducing a 20% initial error. The battery was discharged from 80% SOC to 20% SOC and then charged back to 80% SOC, keeping the operation within the SOC-OCV flat zone at 25°C. More importantly, a 15-minute constant charging current segment (-1 A) was included and is highlighted in Fig. 20.

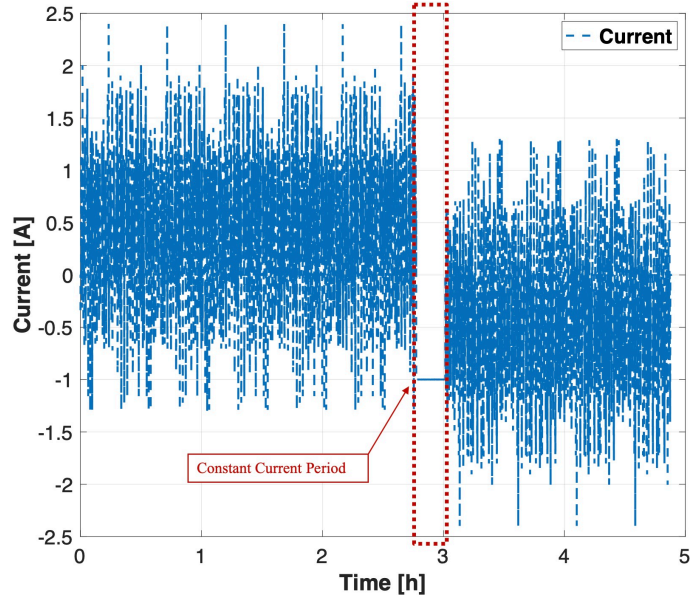

Fig. 20. Current profile with a 15-min constant current segment.

As shown in the lower subplot of Fig. 21, the RMSE of the proposed method is 2.96%. Remarkably, the $SOC_{est}$ matches the true SOC even during the constant current segment. Besides, we can see that once the constant current is applied, leading to a low current excitation level, the value of $SOC_{OCV-H}$ for the Parameter Estimation Module becomes unreliable. However, the value of $cov_{SOC_{OCV-H}}$ also instantaneously increases. Recalling Eq. (20) in the Condition Evaluation Module, when the current excitation level is low, the first element in the inverse Fisher information matrix ($[F(k)^{-1}]_{11}$) increases, leading to a high $cov_{SOC_{OCV-H}}$ value. Thus, with the extremely high $cov_{SOC_{OCV-H}}$ value, the SOC Fusion Module ignores the incorrect $SOC_{OCV-H}$ value due to the constant current and relies purely on Coulomb counting to maintain more reasonable results. This demonstrates how the proposed method exhibits robustness to operating scenarios where the persistence of excitation condition is not satisfied.

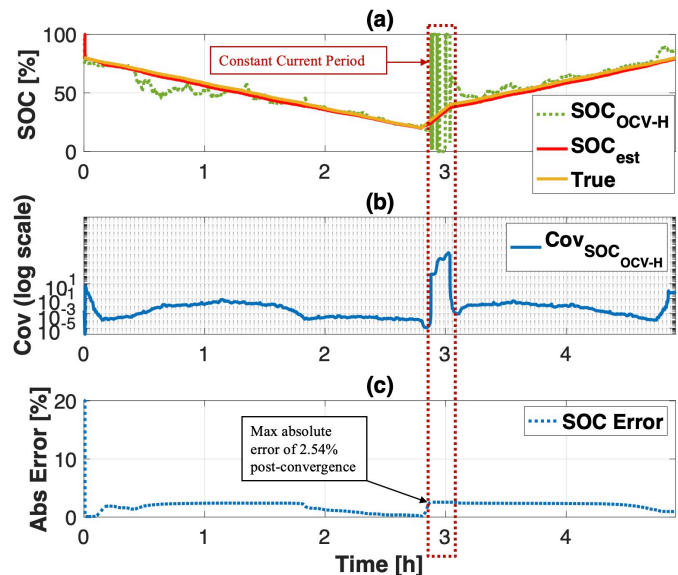

Fig. 21. Proposed method with current profile including a 15-min constant current segment.

## IV. Conclusion and Discussions

### A. Conclusion

This paper presents an adaptive SOC estimation algorithm for LFP batteries that overcomes significant challenges associated with undesirable operational conditions. These challenging operational conditions include prolonged operation in the flat OCV-SOC zone, current measurement bias, low temperature operation, and more. Our proposed system integrates four key modules: the Parameter Estimation Module, the OCV-H-SOC Inversion Module, the Condition Evaluation Module, and the SOC Fusion Module.

The Parameter Estimation Module estimates the model parameters, including OCV. Importantly, it decouples the impact of current bias and OCV estimation. The OCV-H-SOC Inversion Module captures the critical relationship between OCV, hysteresis, and SOC, accounting for hysteresis effects that are particularly crucial for LFP cells. The Condition Evaluation Module utilizes Fisher information and the Cramér-Rao bound to dynamically assess the confidence levels of SOC estimation from OCV inversion, ensuring reliable performance even in SOC-OCV flat zones. Finally, the SOC Fusion Module combines Coulomb counting with voltage-based SOC estimations using a Kalman filter, resulting in a robust and accurate SOC estimate.

Extensive testing under a variety of conditions reveals the performance of the proposed method vs. the unscented Kalman filter (UKF), a state-of-art benchmark algorithm. In ideal conditions, both the proposed method and the UKF achieved good estimation accuracy. However, the proposed method consistently achieved much lower RMSE values across various challenging scenarios, including prolonged operation in SOC-OCV flat zones, current bias, and low temperatures. Notably, the system effectively managed scenarios with insufficient current excitation, maintaining accuracy based on real-time conditions. Interestingly, we found that both methods handled the voltage quantization errors introduced by ADCs effectively because these errors were almost uniformly distributed with a zero-mean value. To further validate the effectiveness of the proposed approach, we conducted a comprehensive benchmarking study with leading data-driven methods, namely, LSTM and a decoder-style Transformer model, both in standalone and hybrid (with Kalman Filter) configurations. The results, summarized in Appendix V.C, show that although deep learning models can capture complex temporal features, their performance deteriorates in out-of-distribution scenarios.

Given that the proposed method updates its parameters in real-time, it has the potential to enhance battery state of health (SOH) estimation. Future work could explore integrating the proposed system with an SOH estimation system [52]. Additionally, the proposed system has a limitation in its ability to overcome the impact of voltage measurement bias, which could be another area for future research. Lastly, as cybersecurity concerns in energy systems [53] gain increasing attention, particularly with the adoption of smart meters and the expanding EV market, improving cybersecurity for battery management systems (BMS) could be a crucial direction for further study.

### B. Discussions

The proposed SOC estimation method significantly enhances SOC accuracy under challenging conditions, such as flat SOC–OCV zones and current sensor bias. By maintaining reliable SOC expectations, it improves range estimation and prevents overcharging or deep discharging, extending battery life and ensuring safety. Unlike traditional methods that require frequent recalibration, this adaptive approach continuously updates its model parameters based on condition evaluation module, reducing maintenance costs and operation downtime. Its ability to self-correct in real-time ensures that the BMS remains accurate despite temperature, load, or battery aging, providing consistent and trustworthy SOC information for optimal energy management [54] and user confidence [55].

Beyond EV use, accurate SOC tracking is crucial for optimizing energy dispatch and maintaining grid stability in large-scale storage systems. The proposed method's robustness under diverse environmental conditions, including temperatures, ensures reliable SOC readings without manual recalibration. By accurately managing SOC, it enables higher utilization efficiency, extending battery lifespan and reducing operational costs. The adaptive nature of the method minimizes maintenance requirements and prevents drift in SOC estimation, making it well-suited for stationary energy storage systems that demand long-term reliability and efficiency.

When compared with the UKF, the proposed method offers better long-term adaptability due to its real-time parameter learning and hybrid fusion design. Additionally, while LSTM and Transformer-based benchmarks show promising performance in standard conditions, they are prone to degradation in scenarios involving sensor bias, quantization, or temperature shifts due to their reliance on historical patterns learned from limited datasets. These data-driven models also lack built-in physical constraints, making them more susceptible to drift and overfitting. The hybrid fusion (LSTM/Transformer + Kalman Filter) mitigates some of these issues, but still falls short of the proposed method in terms of overall accuracy and robustness.

The comparative analysis thus reinforces the merit of combining physics-informed modeling with adaptive and condition-aware fusion strategies. The proposed algorithm provides a balanced, reliable, and computationally efficient solution for real-time SOC estimation in modern battery management systems across diverse applications.

## V. APPENDIX:
### A. Derivation of the Linear-in-Parameter Battery Model

To facilitate real-time parameter estimation, we reformulate the second-order equivalent circuit model of the battery into a linear-in-parameter structure. The original ECM includes an open-circuit voltage (OCV), an ohmic resistance $R_0$, and two RC pairs characterized by $R_1$ and $R_2$, respectively.

The voltage dynamics across each RC branch are given by the differential equations:

$$\dot{V}_1(t) = -\frac{1}{R_1 C_1} V_1(t) + \frac{I(t)}{C_1} \tag{A1.}$$

$$\dot{V}_2(t) = -\frac{1}{R_2 C_2} V_2(t) + \frac{I(t)}{C_2} \tag{A2.}$$

Taking the Laplace transform and solving for $V_1(s)$ and $V_2(s)$:

$$V_1(s) = \frac{R_1}{(R_1 C_1 s + 1)} I(s) \tag{A3.}$$

$$V_2(s) = \frac{R_2}{(R_2 C_2 s + 1)} I(s) \tag{A4.}$$

The terminal voltage in the Laplace domain is:

$$V_T(s) = OCV - R_0 I(s) - V_1(s) - V_2(s) \tag{A5.}$$

Substituting the expressions for $V_1(s)$ and $V_2(s)$:

$$V_T(s) = OCV - R_0 I(s) - \frac{R_1}{(R_1 C_1 s + 1)} I(s) - \frac{R_2}{(R_2 C_2 s + 1)} I(s) \tag{A6.}$$

Where:

$$R_1 C_1 = \tau_1, \quad R_2 C_2 = \tau_2$$

Multiplying both sides by the common denominator $(\tau_1 s + 1)(\tau_2 s + 1)$ yields:

$$\tau_1 \tau_2 s^2 V_T(s) + (\tau_1 + \tau_2) s V_T(s) + V_T(s) = (\tau_1 \tau_2 s^2 + (\tau_1 + \tau_2) s + 1) OCV$$
$$- (\tau_1 \tau_2 s^2 + (\tau_1 + \tau_2) s + 1) R_0 I(s) - R_1 (\tau_2 s + 1) I(s) - R_2 (\tau_1 s + 1) I(s) \tag{A7.}$$

Grouping terms by powers of $s$, the expression becomes:

$$V_T(s) = OCV - as^2 I(s) - bs I(s) - c I(s) - ds^2 V_T(s) - es V_T(s) \tag{A8.}$$

Where the coefficients are defined as:

$$\begin{cases} a = \tau_1 \tau_2 R_0 \\ b = R_0 \tau_1 + R_0 \tau_2 + R_1 \tau_2 + R_2 \tau_1 \\ c = R_0 + R_1 + R_2 \\ d = \tau_1 \tau_2 \\ e = \tau_1 + \tau_2 \end{cases} \tag{A9.}$$

Taking the inverse Laplace transform gives the time-domain representation:

$$V_T(t) = OCV - a\,\ddot{I}(t) - b\,\dot{I}(t) - c\,I(t) - d\,\ddot{V}_T(t) - e\,\dot{V}_T(t) \tag{A10.}$$

This can be rearranged into a linear-in-parameter form:

$$V_T = \begin{bmatrix} OCV & a & b & c & d & e \end{bmatrix} \begin{bmatrix} 1 \\ -\ddot{I} \\ -\dot{I} \\ -I \\ -\ddot{V}_T \\ -\dot{V}_T \end{bmatrix} \tag{A11.}$$

This linear form enables the use of standard linear regression techniques for real-time parameter identification. The voltage and current derivatives are computed using filtered signals as described in Appendix B.

*B. Derivation of Filtered and Discretized Signal Derivatives*

To support real-time parameter estimation, we extract filtered signals and their time derivatives from raw voltage and current measurements using a second-order filtering process. To estimate the filtered voltage and current signals and their time derivatives, we apply zero-order hold (ZOH) discretization to the continuous-time transfer functions $G_0(s)$, $G_1(s)$, and $G_2(s)$. These transfer functions are used to extract the signal itself, its first derivative, and its second derivative, respectively.

After ZOH discretization, we obtain the discrete-time transfer functions $G_0(z)$, $G_1(z)$, and $G_2(z)$ in the z-domain as:

$$G_2(z) = \frac{\lambda_0(z-1)^2}{T_s^2 z^2 \left(\lambda_0 + \frac{\lambda_1(z-1)}{T_s z} + \frac{(z-1)^2}{T_s^2 z^2}\right)} \tag{A12.}$$

$$G_1(z) = \frac{\lambda_0(z-1)}{T_s z \left(\lambda_0 + \frac{\lambda_1(z-1)}{T_s z} + \frac{(z-1)^2}{T_s^2 z^2}\right)} \tag{A13.}$$

$$G_0(z) = \frac{\lambda_0}{\lambda_0 + \frac{\lambda_1(z-1)}{T_s z} + \frac{(z-1)^2}{T_s^2 z^2}} \tag{A14.}$$

Each transfer function $G_i(z)$ (for $i = 0,1,2$) is then converted into a discrete-time state-space representation with system matrices $A_i, B_i, C_i, D_i$.

Using the state-space form, the filtered terminal voltage $\hat{V}_T[k]$, current $\hat{I}[k]$, and their first and second derivatives are calculated as:

Filtered Voltage:

$$x_{V_T}[k+1] = A_0 x_{V_T}(k) + B_0 V_T(k) \tag{A15.}$$

$$\hat{V}_T[k] = C_0 x_{V_T}(k) + D_0 V_T(k) \tag{A16.}$$

First Derivative of Voltage:

$$x_{\dot{V}_T}[k+1] = A_1 x_{\dot{V}_T}(k) + B_1 V_T(k) \tag{A17.}$$

$$\hat{V}'_T[k] = C_1 x_{\dot{V}_T}(k) + D_1 V_T(k) \tag{A18.}$$

Second Derivative of Voltage:

$$x_{\ddot{V}_T}[k+1] = A_2 x_{\ddot{V}_T}(k) + B_2 V_T(k) \quad (A19.)$$

$$\hat{V}_T''[k] = C_2 x_{\ddot{V}_T}(k) + D_2 V_T(k) \quad (A20.)$$

Filtered Current:

$$x_I[k+1] = A_0 x_I(k) + B_0 I(k) \quad (A21.)$$

$$\hat{I}[k] = C_0 x_I(k) + D_0 I(k) \quad (A22.)$$

First Derivative of Current:

$$x_{\dot{I}}[k+1] = A_1 x_{\dot{I}}(k) + B_1 I(k) \quad (A23.)$$

$$\hat{I}'[k] = C_1 x_{\dot{I}}(k) + D_1 I(k) \quad (A24.)$$

Second Derivative of Current:

$$x_{\ddot{I}}[k+1] = A_2 x_{\ddot{I}}(k) + B_2 I(k) \quad (A25.)$$

$$\hat{I}''[k] = C_2 x_{\ddot{I}}(k) + D_2 I(k) \quad (A26.)$$

*C. Benchmark Comparison with Data-Driven Models (LSTM and Transformer)*
Appendix C provides a comprehensive benchmarking study comparing the proposed method with leading data-driven approaches, LSTM and Transformer models, both standalone and fused with a Kalman Filter. This section details the network architectures, fusion strategies, test conditions, and computational costs, and presents comparative results under a range of realistic operating scenarios. The analysis highlights the strengths and limitations of each approach, offering insights into their suitability for real-time LFP SOC estimation.

To establish a robust data-driven benchmark for SOC estimation, we implemented a deep learning model based on Long Short-Term Memory (LSTM) networks. The model was trained using real driving data collected from multiple battery test cycles. To ensure generalizability, all available valid cycles were used for training. A comprehensive set of features was engineered, including voltage, current, their derivatives, moving averages, and instantaneous power, followed by z-score normalization. A sequence length of 120 with a stride of 8 was chosen to balance temporal resolution and computational efficiency. The architecture consisted of a bidirectional LSTM layer with 256 hidden units to capture temporal dependencies, followed by a unidirectional LSTM layer with 128 units, and a series of dense layers with dropout and batch normalization for regularization and improved convergence. The network was trained using the Adam optimizer and adaptive learning rate scheduling. This LSTM-based model provides a strong deep learning baseline that reflects the capability of sequence modeling architectures in learning SOC behavior from raw voltage and current data alone.

To enhance the robustness and accuracy of SOC estimation, we implemented a sensor fusion strategy that combines Coulomb Counting (CC) with LSTM-based predictions using a one-dimensional Kalman Filter (KF). The system dynamics model assumes SOC evolution based on CC, accounting for coulombic efficiency during charging and a small self-discharge rate, while treating the LSTM output as a noisy measurement of the true SOC. Specifically, the prediction step uses the current and battery capacity to estimate SOC incrementally, introducing process noise to capture uncertainties in CC. The correction step incorporates the LSTM-based SOC estimate when available, adjusting the state estimate based on the Kalman gain computed from process and measurement noise covariances. The filter was initialized with a small state covariance and tuned with empirically selected values for process (Q = $10^{-7}$) and measurement (R = $10^{-2}$) noise. The LSTM measurements are fused starting from the earliest index where sequence input becomes available. This fusion approach leverages the high-frequency consistency of CC and the learned nonlinear modeling capacity of LSTM, resulting in a more stable and accurate SOC estimate that mitigates the weaknesses of each individual source. Physical bounds were enforced throughout to ensure realistic SOC values.

We also implemented a decoder-only Transformer model with masked multi-head self-attention, tailored for real-time SOC prediction. This architecture follows the canonical Transformer decoder structure, leveraging stacked layers of masked multi-head self-attention and position-wise feedforward networks, along with residual connections and layer normalization, to capture long-range temporal dependencies in the input sequences. To ensure real-time compatibility and strict temporal causality, masked self-attention is applied in each decoder block so that the model only attends to historical and current measurements (voltage and current), never using any future information for prediction. The input to the network is a two-channel sequence (voltage and current) with a length of 120 timesteps, corresponding to a 2-minute history at 1 Hz sampling. These inputs are standardized using z-score normalization with normalization parameters recorded for deployment compatibility. The model consists of six stacked decoder blocks, each composed of masked multi-head self-attention (with 8 attention heads and model dimension of 192), feedforward layers (hidden size 384 followed by 192), dropout layers (rates ranging from 2% to 6%), and residual skip connections with layer normalization. The final output is generated by applying a fully connected regression head to the representation at the last timestep. The network was trained using the Adam optimizer with a learning rate of 0.0001, a mini-batch size of 12, a total of 60 epochs, and L2 regularization of $3\times10^{-6}$. Validation was performed using an 80/20 split of the sequences. This model provides a strong ability to capture complex nonlinear dynamics and hysteresis behavior in LFP batteries. While recurrent neural networks such as LSTMs have been more commonly used for SOC estimation, Transformer-based causal sequence models are a recent and promising approach in this domain, making this architecture a valuable and rigorous benchmark for evaluating the performance of real-time data-driven SOC estimation methods under challenging operating conditions.

To further enhance the robustness of the Transformer-based SOC predictions and reduce the inherent short-term noise observed in purely data-driven outputs, we developed a KF-based fusion approach similar to the LSTM + KF method. Here, SOC evolution is again propagated using CC, and the Transformer output is incorporated as a noisy measurement. The same KF parameters were used ($Q = 10^{-7}$, $R = 10^{-2}$, $P_0 = 10^{-2}$), and Transformer predictions are fused when available. This hybrid Transformer + KF model combines the Transformer's strong nonlinear modeling capabilities with the temporal consistency and stability of CC, producing smoother and more reliable SOC estimates while maintaining real-time compatibility.

To validate the effectiveness of these methods, we conducted a comparative analysis summarized in Table A.1 and Fig. A1 through A5. In addition to the data-driven methods, we included the unscented Kalman filter (UKF) as a reference baseline. While the UKF's structure and performance have been discussed in detail in the main body of the paper, we include its results here for completeness and comparative context.

Table A.1 presents the root mean square error (RMSE) of SOC estimation under five representative conditions: ideal case, SOC-OCV flat zone, current bias, quantization error, and low temperature. Across all cases, the proposed method consistently outperformed the benchmarks. Both Transformer + KF and LSTM + KF showed notable improvements over their standalone counterparts, validating the effectiveness of hybrid fusion.

Table A1. SOC Estimation RMSE (%) Across Different Test Conditions

| Test Condition | Proposed | UKF | LSTM | LSTM + KF | Transformer | Transformer + KF |
|---|---|---|---|---|---|---|
| Ideal Case | 0.49% | 3.92% | 5.51% | 4.32% | 3.197% | 2.36% |
| SOC-OCV Flat Zone | 2.54% | 6.69% | 7.24% | 5.97% | 7.02% | 5.52% |
| Constant Current Bias | 2.99% | 15.44% | 8.19% | 6.77% | 9.58% | 5.86% |
| Quantization Error | 2.69% | 7.13% | 7.27% | 5.99% | 7.056% | 5.59% |
| Low Temperature | 3.28% | 28.11% | 11.42% | 8.69% | 9.95% | 8.29% |

Table A.2 reports the computational cost of each method per prediction. All computations were performed on a standard laptop (Apple M1 Max, 64 GB memory). Execution times were averaged over 129,618 steps (equivalent to 36 hours of operation at a 1-second sampling interval) and normalized per time step. As expected, the UKF achieved the lowest cost (0.63 ms) due to its lightweight structure. The LSTM- and Transformer-based models required more computation (1.31 ms and 1.78 ms, respectively) owing to their deeper neural network architectures. Notably, augmenting these models with a Kalman Filter introduced only minimal additional overhead, approximately 0.005 ms, while substantially improving robustness and accuracy under challenging test conditions. The proposed method, which incorporates interpolation and regression modules, achieved a balanced runtime of 1.17 ms. All methods remain well within acceptable limits for real-time onboard SOC estimation, and the fusion strategies (LSTM + KF and Transformer + KF) represent a compelling trade-off between performance and computational efficiency.

Table A2. Computational Cost per Prediction

| | Proposed | UKF | LSTM | LSTM + KF | Transformer | Transformer + KF |
|---|---|---|---|---|---|---|
| **Operation time** | 1.17 ms | 0.63 ms | 1.31 ms | 1.32 ms | 1.78 ms | 1.79 ms |

Under the ideal test condition (Fig. A1), all models achieved relatively low errors. The proposed method yielded the best performance, followed by Transformer + KF (2.36%). The Transformer's ability to model long-term patterns and the KF's grounding in physical dynamics created a complementary effect. Notably, Transformer outperformed LSTM even as a standalone method, highlighting its strength in capturing global sequence structures.

In the SOC-OCV flat zone (Fig. A2), the estimation challenge increased due to poor observability from voltage signals—a characteristic limitation of LFP chemistry. All models experienced degraded performance, particularly the UKF, which relies on voltage for state correction. However, the fusion-based models (Transformer + KF and LSTM + KF) maintained better accuracy by leveraging current integration from Coulomb Counting (CC) to supplement the limited information from voltage.

In the constant current bias scenario (Fig. A3), the UKF performed poorly due to its dependence on direct current measurements and the resultant accumulation of integration errors under bias. Purely data-driven models such as Transformer and LSTM were more robust to current bias, owing to their learned sequence representations, but still exhibited prediction drift in the absence of physical correction. The fusion approaches also degraded in this setting, as the integrated CC model was itself affected by the bias-induced drift, limiting the effectiveness of the Kalman correction.

For quantization error (Fig. A4), fusion again proved beneficial. While voltage quantization reduced the resolution of observable dynamics, CC-based models smoothed over the noise, and LSTM + KF and Transformer + KF achieved robust performance (5.99% and 5.59%, respectively). This case illustrates that voltage quantization errors introduced by ADCs can be mitigated if appropriately filtered in all methods.

At low temperatures (Fig. A5), all methods deteriorated, with UKF showing the largest error (28.11%) due to unmodeled temperature-dependent behaviors. Data-driven models also degraded, highlighting their sensitivity to out-of-distribution conditions. Fusion strategies showed better generalization (8.69 % and 8.29% for LSTM + KF and Transformer + KF, respectively) but still lagged behind the proposed method (3.28%).

In summary, while deep learning models such as LSTM and Transformer can capture complex, nonlinear battery behaviors and achieve high accuracy on representative datasets, they struggle to generalize under distribution shifts, sensor noise, or operating conditions not seen during training. Their lack of physical constraints can lead to prediction drift and unrealistic SOC estimates. By fusing these models with physically grounded CC via a Kalman Filter, both robustness and interpretability are improved, resulting in more stable and physically plausible predictions. Although these hybrid approaches still trail the proposed physics-informed method in overall performance, they offer strong, real-time-capable benchmarks for SOC estimation. Future work could explore incorporating additional physical variables, such as temperature, or applying domain adaptation and online recalibration to further enhance the generalization ability of data-driven models.

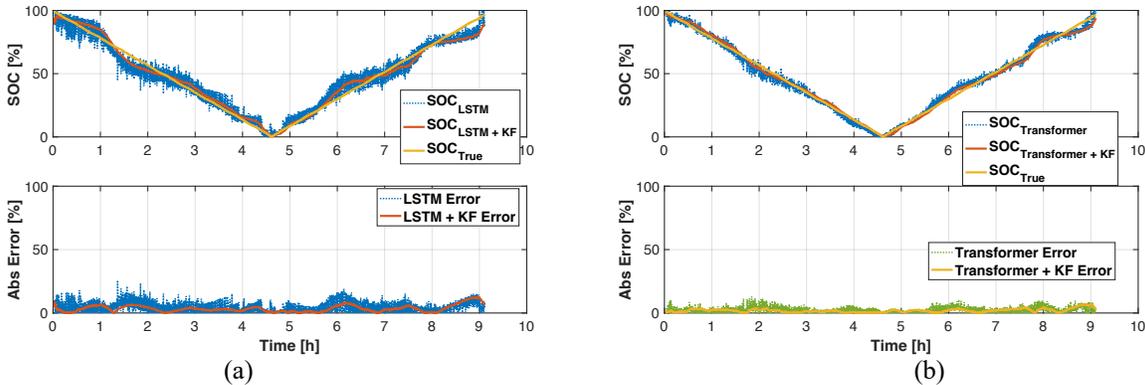

Fig. A1. Test results under ideal condition. (a) LSTM & LSTM + Kalman filter. (b) Transformer & Transformer + Kalman filter.

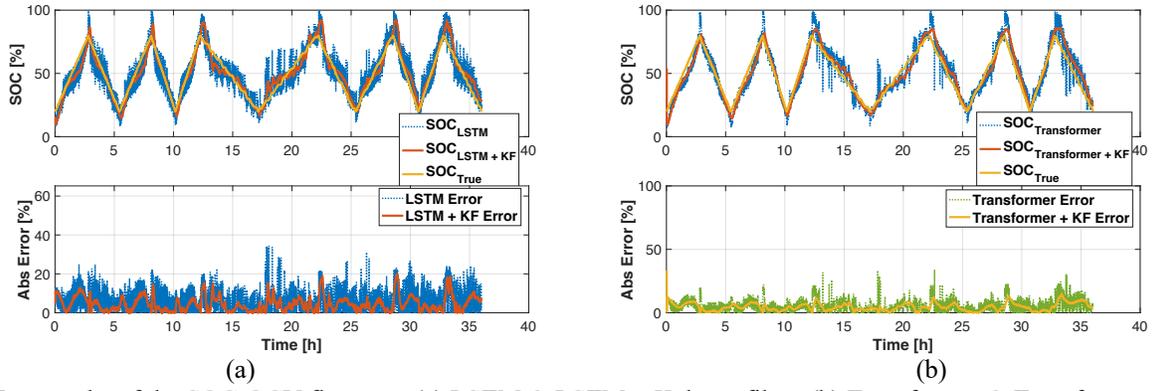

Fig. A2. Test results of the SOC-OCV flat zone. (a) LSTM & LSTM + Kalman filter. (b) Transformer & Transformer + Kalman filter.

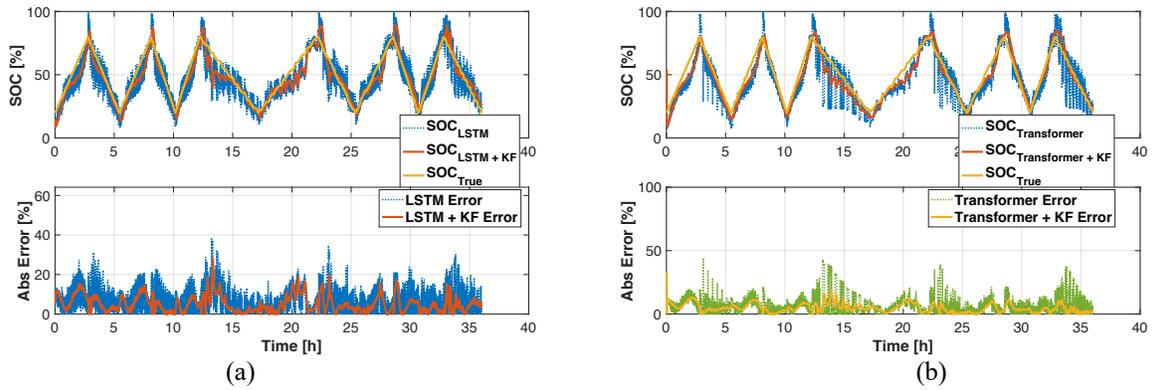

Fig. A3. Test results of the SOC-OCV flat zone adding a high current measurement bias (-0.05A). (a) LSTM & LSTM + Kalman filter. (b) Transformer & Transformer + Kalman filter.

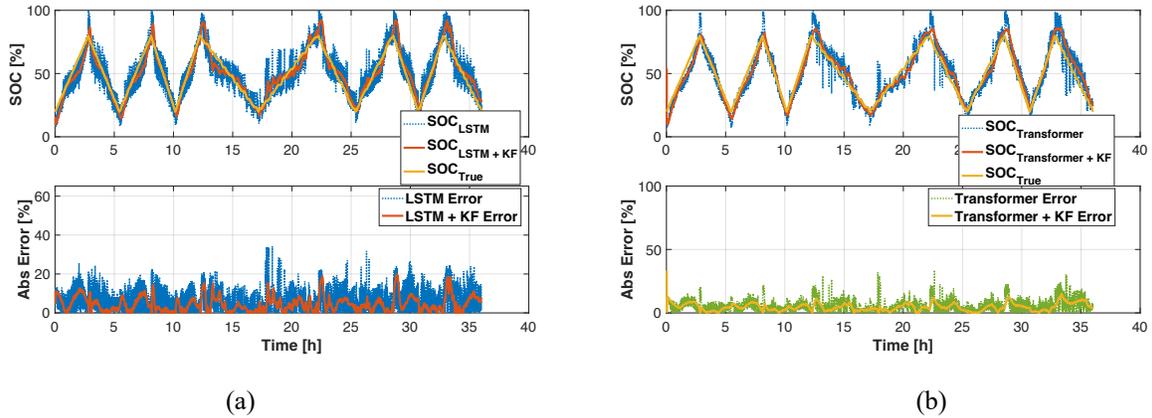

Fig. A4. Test results of the SOC-OCV flat zone with voltage quantization errors (10-bit ADC). (a) LSTM & LSTM + Kalman filter. (b) Transformer & Transformer + Kalman filter.

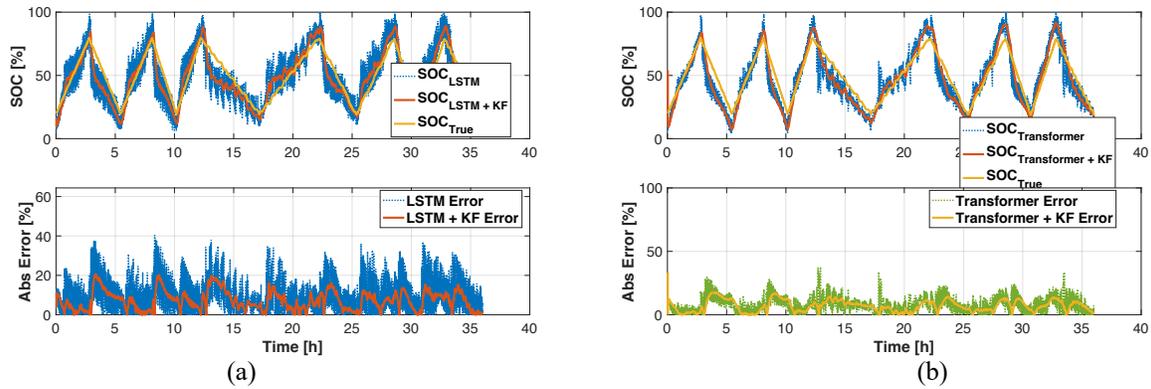

Fig. A5. Test results of the SOC-OCV flat zone and low temperature (10°C). (a) LSTM & LSTM + Kalman filter. (b) Transformer & Transformer + Kalman filter.